\documentclass[iop]{emulateapj}

\usepackage[usenames, dvipsnames]{color}

\newcommand{\myemail}{matthew@astro.su.se}

\slugcomment{Accepted by the Astrophysical Journal}

\shorttitle{O{\sc vi} Mapping of a Starburst Galaxy}
\shortauthors{Hayes et al.}

\newcommand{\lya}{\mbox{Ly$\alpha$}}
\newcommand{\lyb}{\mbox{Ly$\beta$}}

\newcommand{\halpha}{\mbox{H$\alpha$}}
\newcommand{\hbeta}{\mbox{H$\beta$}}

\newcommand{\hI}{\mbox{H~{\sc i}}}
\newcommand{\hII}{\mbox{H~{\sc ii}}}

\newcommand{\heII}{\mbox{He~{\sc ii}}}

\newcommand{\oI}{\mbox{O~{\sc i}}}
\newcommand{\oII}{\mbox{O~{\sc ii}}}
\newcommand{\oIII}{\mbox{O~{\sc iii}}}
\newcommand{\oVI}{\mbox{\normalfont{O}~{\sc vi}}}
\newcommand{\oVII}{\mbox{O~{\sc vii}}}
\newcommand{\cII}{\mbox{C~{\sc ii}}}
\newcommand{\sII}{\mbox{S~{\sc ii}}}

\newcommand{\NoVI}{\mbox{$N_\mathrm{OVI}$}}

\newcommand{\kms}{\mbox{km\,s$^{-1}$}}
\newcommand{\ergsec}{\mbox{erg~s$^{-1}$}}

\newcommand{\ergseccmarcsec}{\mbox{erg~s$^{-1}$~cm$^{-2}$~arcsec$^{-2}$}}

\newcommand{\ergseccm}{\mbox{\mbox{erg~s$^{-1}$~cm$^{-2}$}}}
\newcommand{\msun}{\mbox{M$_\odot$}}

\newcommand{\msunyr}{\mbox{M$_\odot$~yr$^{-1}$}}
\newcommand{\msunyrkpc}{\mbox{M$_\odot$~yr$^{-1}$~kpc$^{-2}$}}

\newcommand{\ebv}{\mbox{$E_{B-V}$}}

\newcommand{\percm}{\mbox{cm$^{-2}$}}

\newcommand{\mstell}{\mbox{$M_\star$}}

\newcommand{\lha}{\mbox{$L_{\mathrm{H}\alpha}$}}

\newcommand{\ewha}{\mbox{$W_{\mathrm{H}\alpha}$}}

\newcommand{\izw}{\mbox{{\sc i}\,Zw\,18}}
\newcommand{\sbs}{\mbox{SBS\,0335--052}}
\newcommand{\har}{\mbox{Haro\,11}}

\begin{document}

\title{\oVI\ Emission Imaging of a Galaxy with the Hubble Space Telescope: a
Warm Gas Halo Surrounding the Intense Starburst SDSS\,J115630.63+500822.1}

\author{Matthew Hayes\altaffilmark{2,3}}
\author{Jens Melinder\altaffilmark{3}}
\author{G\"oran \"Ostlin\altaffilmark{3}}
\author{Claudia Scarlata\altaffilmark{4}}
\author{Matthew D. Lehnert\altaffilmark{5}}
\author{Gustav Mannerstr\"om-Jansson\altaffilmark{3}}

\altaffiltext{1}{Based on observations made with the NASA/ESA Hubble Space
Telescope, obtained at the Space Telescope Science Institute, which is operated
by the Association of Universities for Research in Astronomy, Inc., under NASA
contract NAS 5-26555. These observations are associated with program \#13656.}
\altaffiltext{2}{\myemail}
\altaffiltext{3}{Stockholm University, Department of Astronomy and Oskar Klein
Centre for Cosmoparticle Physics, AlbaNova University Centre, SE-10691,
Stockholm, Sweden.}
\altaffiltext{4}{Minnesota Institute for Astrophysics, School of Physics and
Astronomy, University of Minnesota, 316 Church Str. SE, Minneapolis, MN 55455,
USA.}
\altaffiltext{5}{Institut d'Astrophysique de Paris, UMR 7095, CNRS, Universit\'e
Pierre et Marie Curie, 98bis boulevard Arago, 75014, Paris, France.}

\begin{abstract}
We report results from a new HST campaign that targets the \oVI~$\lambda\lambda
1032,1038$~\AA\ doublet in emission around intensely star-forming galaxies.  The
programme aims to characterize the energy balance in starburst galaxies and gas
cooling in the difficult-to-map coronal temperature regime of $2-5\times
10^5$~Kelvin.  We present the first resolved image of gas emission in the \oVI\
line.  Our target, SDSS\,J115630.63+500822.1, is very compact in the continuum
but displays \oVI\ emission to radii of 23~kpc.  The surface brightness profile
is well fit by an exponential with a scale length of 7.5~kpc.  This is ten times
the size of the photoionized gas, and we estimate that about 1/6 the
total \oVI\ luminosity comes from resonantly scattered continuum radiation. 
Spectroscopy -- which closely resembles a stacked sample of archival spectra --
confirms the \oVI\ emission, and determines the column density and outflow
velocity from blueshifted absorption.  The combination of measurements enables a
large number of calculations with few assumptions.  The \oVI\ regions fill only
$\sim 10^{-3}$ of the volume.  By comparing the cooling time with the cloud
sound-crossing time, the cooling distance with the size, and the pressure in the
\oVI\ and nebular gas, we conclude that the \oVI-bearing gas cannot have been
lifted to the scale height at this temperature, and must be cooling in situ
through this coronal temperature regime.  The coronal phase contains $\sim 1$~\%
of the ionized mass, and its kinetic energy  at a given instant is $\sim 1$~\%
of the budget set by supernova feedback.  However a much larger amount of the
gas must have cooled through this phase during the star formation episode.  The
outflow exceeds the escape velocity and the gas may become unbound, but it will
recombine before it escapes and become visible to Lyman (and \oI) spectroscopy.  
The mapping of this gas represents a crucial step in further constraining galaxy
formation scenarios and guiding the development of future astronomical
satellites.
\end{abstract}


\section{Introduction}

Gaseous outflows from galaxies, taking the form of large-scale `superwinds',
become commonplace once a certain threshold surface density of star-formation is
exceeded \citep[$\Sigma_\mathrm{SFR} \gtrsim
0.1$~\msunyrkpc;][]{Heckman2002conf}.  For local galaxies considered starbursts,
outflows of high- and low-ionization gas are effectively ubiquitous
\citep{Heckman1990,Rupke2005,Martin2005,Heckman2011,Rivera-Thorsen2015}, and
comparable results are also found at higher redshifts
\citep[e.g.][]{Weiner2009,Rubin2010,Erb2012,Rubin2014}, including a similar
threshold surface density of star-formation \citep{Kornei2012}.  

The most likely mechanism for driving large-scale outflows begins with the
ejection of fast-moving material in the winds of the most massive stars, and the
ejecta of core-collapse supernovae.  In dense star forming regions these ejecta
collide, the gas shock heats and thermalizes, and produces a hot plasma in the
nuclear regions.  With temperatures of around $5\times10^7$~K -- seen
observationally using iron lines in the hard X-ray bands
\citep[e.g.][]{Strickland2009} -- this gas rapidly expands and flows outwards
from the (central, likely nuclear) star-forming regions, imparting energy and
momentum onto surrounding media, and sweeping up material in various phases
\citep{Chevalier1985,Murray2005}.  Due to the hydrodynamical interactions
between hot and cold gas phases, and compression, heating and enhanced cooling
at the interfaces, such outflows are genuine multiphase phenomena
\citep{McKee1977,Hopkins2012,Braun2012,Thompson2016}.  Indeed outflow signatures
are observed in almost all phases from starbursts and active galactic nuclei
\citep{Grimes2009,Rivera-Thorsen2015,Feruglio2015,Chisholm2015,Beirao2015}. 

This feedback from star formation is probably a vital ingredient in the shaping
of a number of galaxy scaling relations.  Feedback is believed to be the main
source for regulating the subsequent star formation in galaxies, preventing too
many stars from being formed too early.  In balancing the rate at which gas is
accreted onto halos, it is responsible for keeping galaxies on the approximately
linear relationship between star formation rate (SFR) and stellar mass
(\mstell): the `main sequence'.  Consequently, stellar feedback and its
efficiency are likely responsible for shaping at least the low luminosity end of
the galaxy luminosity function \citep[e.g.][]{Efstathiou2000,Benson2003}.
Moreover, since outflows are driven in part by supernova ejecta, they are likely
to be metal-enriched compared to the interstellar medium (ISM).  If winds escape
from galaxies then feedback will also impact the shaping of the relationship
between galaxy metallicity and mass \citep[e.g.][]{Tremonti2004,Andrews2013}. 

Since galaxy formation simulations are generally regarded as successful if they
can reproduce the above scaling relations \citep[see][ for a
review]{Somerville2015}, it must be vital that the prescriptions of feedback are
as precise as possible.  Not only must one know whether winds will escape from
the halos of the galaxies that launch them (thereby enriching the intergalactic
medium, IGM, with metals), but also what phase the gas will be in if and when it
does.  I.e. if it will remain in a hot low-density phase, in which case it
cannot contribute to subsequent star-formation anywhere, or if it will cool,
recombine, and perhaps re-accrete, and rejoin the reservoir of available
material.  From the observational perspective this requires us to determine the
spatial distribution and kinematics of gas in the ISM and circumgalactic medium
(CGM).  Since the outflows are likely multi-phase, existing over a wide range of
temperatures that account for different amounts of the energy and mass, ideally
one would do this with as many tracers as possible, from the hottest gas that
drives that initial stages of the outflow, to the cold gas phases that are
entrained.  

The hottest gas -- from roughly a million K and upwards -- may be traced by
X-ray emission, while from temperatures of a few times $10^4$~K observers
utilize nebular line emission in the optical, transitions of low ionization
stage metals in the far ultraviolet, and molecular lines in the radio (in order
of decreasing temperature).  This leaves a gap of between a few $10^4$ and
$10^6$~K, that can only be probed using the resonant transitions of highly
ionized metals in the far ultraviolet.  At these temperatures, of
several $10^5$~K the strongest and most readily available is the doublet of
\oVI\ at $\lambda\lambda=1032,1038$~\AA\ in the restframe.   Such species
remain vital, being the main coolants of metal enriched gas at these
temperatures \citep[e.g.][]{Sutherland1993}, and thus its availability and
distribution governs the nature and evolution of gas in galaxy halos.  In this
paper we address this phase directly, by measuring the morphology of $T \approx
300,000$~K gas in the halo of a strongly star-forming galaxy with the Hubble
Space Telescope (HST).

It is hard to understate the relevance of \oVI\ in this context.  Oxygen is the
third most abundant element in the Universe, is produced by stars, and thus
should accurately trace the growth of metals created through stellar
nucleosynthesis, the majority of which is ejected into the ISM.  
However, when observers attempt to account for the metals in galaxies, a
significant deficit is usually found; this holds for a variety of galaxy
selection functions and over a wide range of redshifts
\citep{Prochaska2003,Bouche2005,Zahid2012}.  Focusing specifically on the metals
traced by \oVI, recent studies with the \emph{Cosmic Origins Spectrograph} (COS)
on HST have proven pivotal in unveiling the reservoir of warm oxygen, as the
metals are expelled to large galactocentric radii.  Specifically the
\emph{COS-Halos} survey \citep{Tumlinson2011sci,Tumlinson2013} has precisely
quantified the amount of \oVI\ gas in galaxy halos: summing the oxygen found in
stars, dust, and various gas phases, \citet{Peeples2014} find that only $\approx
30$~\% on average of the metals produced by stars can be accounted for.
Particularly in lower mass galaxies, $\mstell\approx 10^{9.5}$~\msun, the bulk
of this oxygen resides in a diffuse phase, in circumgalactic environments at
distances of up to $\sim 150$~kpc \citep[see also][]{Prochaska2011ovi}.  Only
for these lower mass galaxies, and under the most optimistic of assumptions, can
the metal budget be recovered. 

The identification of warm metals has been a major avenue of exploration with
mid and high resolution ultraviolet spectrographs, such as the \emph{Far
Ultraviolet Spectroscopic Explorer} (FUSE) and the \emph{Space Telescope Imaging
Spectrograph} (STIS) and COS on HST
\citep{Tripp2000,Hoopes2002,Sembach2004,Tumlinson2005,Prochaska2006,Danforth2006,Tripp2008,
Thom2008,Wakker2009,Savage2011,Tumlinson2013,Stocke2013, Savage2014}.  By
targeting metal enriched gas at $T\sim 10^{5.5}$~K, these studies are intimately
linked with the more general problem of absent baryonic matter at low-$z$ -- the
`missing baryons problem' \citep[see][for a review]{Bregman2007}.  While
enormously successful, absorption line studies rely upon the presence a
UV-bright background quasar along the line-of-sight, and two limitations are
implicit in the method.  Firstly appropriately bright QSOs are rare, and not any
object can be observed; while disks that form stars may be abundant at low-$z$,
extreme starbursts are not, which means appropriate QSO-starburst projections
will be particularly rare.  Such starbursts may be enormously important because
these are the objects that drive the strongest winds at low-$z$
\citep{Heckman2011,Heckman2015}.  Moreover, they share many of their properties
with galaxies at $z>1$ where the bulk of the cosmic metals were produced, and
indeed metal enrichment in the universe must have begun early
\citep[e.g.][]{Cen2006}.  The second drawback is that unless the galaxy is very
extended on the sky, the average galaxy will not have a sufficiently bright
background QSO and, if it does, a measurement will be possible
(probabilistically) on only one sightline.  Thus galaxies can only be studied
population-wise, by combining single sightlines through many objects. 

Both of the above issues could be overcome if we could map the \oVI\ gas in
emission, and infer column and volume densities from the surface brightness.
Indeed detecting the warm-hot intergalactic medium via \oVI\ emission is one of
the goals of the \emph{Faint Intergalactic Redshifted Emission Balloon}
\citep[FIREBALL,][]{Tuttle2010} and \emph{Advanced Technology Large Aperture
Space Telescope} \citep[ATLAST,][]{Postman2010}, albeit with more explicit focus
on extended intergalactic gas and the large-scale structure of the cosmic web.
With current facilities the detection remains a particularly challenging task,
because the densities in the coronal \oVI\ phase are low, and the
$n_\mathrm{e}^2$ dependence of the emissivity suggests that the surface
brightness would also be low.  The \oVI$\lambda\lambda 1032,1038$~\AA\ doublet
has, however, been detected in emission in a small number of cases: NGC\,4631
\citep{Otte2003} and \har\ \citep{Grimes2007} both exhibit irrefutable \oVI\
emission lines which lends some confidence to the prospect of mapping the gas.  

A second challenge is that to map the emission line, a far ultraviolet
narrowband filter or integral field spectrograph would be needed, and no such
dedicated hardware is currently available.  However over the last decade our
group has developed methods of synthesizing narrowband filters using existing
channels of the \emph{Solar Blind Channel} (SBC) of the \emph{Advanced Camera
for Surveys} (ACS) on HST, which we have verified many times by targeting the
\lya\ line of \hI\ \citep{Hayes2005,Hayes2009,Hayes2013}.  

In this paper we present the first \oVI\ emission-line imaging observations of a
star-forming galaxy: SDSS\,J115630.63+500822.1.  The combination of HST emission
imaging and absorption spectroscopy allows us to calculate a large number of
properties relevant for the warm coronal gas phase, how it is excited, and what
its future will be.  As we present only one single galaxy, this stage of the
project can be regarded as a pilot study, although it will be expanded in the
current observing cycle of HST.  We have selected an intense local starburst
that hitherto has not been studied in much detail, although it does enter the
sample of highly excited blue compact galaxies studied by \citet{Chung2013}.  We
present the motivation for selecting this target in Section~\ref{sect:target}
and derive a number of common physical properties from the optical spectrum and
spectral energy distribution.  The method used to isolate the \oVI\ line is
presented in Section~\ref{sect:obs}, together with a description of the HST
observations.  Section~\ref{sect:reduct} presents the data reduction and
processing.  In Section~\ref{sect:res} we present the results, that show a large
extended halo of excess emission in the on-line bandpass and a faint emission
line in the spectrum that we interpret as \oVI$\lambda 1038$~\AA.  In
Section~\ref{sect:archive} we present an analysis of archival COS spectra of
starbursts at similar redshifts; we perform a stacking analysis to show that the
same spectral emission feature is also visible in other galaxies.
Section~\ref{sect:interp} presents a set of calculations in which we estimate
many properties of warm gas.  These include the size of the emitting regions,
the mass in this phase, the cooling time and the ultimate fate of the gas.
These are the main results of the Paper.  In Section~\ref{sect:lit} we compare
the observational results obtained here with the other two detections of \oVI\
emission from galaxies.  Finally in Section~\ref{sect:conc} we present a summary
and our conclusions.  We adopt cosmological parameters of
$H_0=70$~km~s$^{-1}$~Mpc$^{-1}$, $\Omega_\mathrm{M}=0.3$, and
$\Omega_\Lambda=0.7$.   Any magnitudes will be given in the AB system
\citep{Oke1983}.

\section{The Target Galaxy: J1156+5008}\label{sect:target}

The target galaxy has never been discussed in detail before, so we here present
the method and motivations we used to select it, and some of its basic
(`standard') measured and derived properties. 

\subsection{Motivations and Selection}

Only two individual star-forming galaxies are known to emit \oVI: NGC\,4631
\citep{Otte2003} and \har\ \citep{Bergvall2006,Grimes2007}. Prior to our study
we had no information that we could use to select targets.  Moreover, both of
these previous detections were in spectroscopic observations, made with the
FUSE.  However the FUSE had a large ($30\times 30$ arcsec) square aperture and
provided no spatial information that could guide us -- the only `morphological'
information available was that \oVI\ emission is observed away from the disk in
NGC\,4631.  Without knowledge of what galaxies we should target or where best to
observe them, we selected the star-forming galaxy most likely to emit \oVI\
based mainly upon results from the FUSE archive and the \emph{COS-Halos} survey
\citep{Tumlinson2013}.  

\citet{Tumlinson2011sci} demonstrated that \oVI\ absorption was ubiquitous in
the halos of star-forming galaxies but not necessarily present in passive
systems.  Within their star-forming sample, they identified a trend in which the
\oVI\ column density (\NoVI) increases with increasing sSFR.  While this does
not tell us that the volume density (the important quantity for emissivity) also
increases, we argued that a correlation would most likely be positive.  However
when studying \oVI\ absorption against the UV continuum of galaxies themselves,
the ubiquity of \NoVI\ features does not extend to the low-mass dwarf
starbursts: \izw\ and \sbs\ do not show \oVI\ absorption in FUSE spectra
\citep{Grimes2009}.  We suspect the reasons for this are are firstly that these
galaxies are extremely metal poor ($\sim1$/40 the solar value in most metallic
ions), and secondly that the starbursts are particularly young and they have not
yet developed large-scale supernova-driven outflows to collisionally excite the
warm phase.  Since the winds and chemical enrichment are primarily the result of
supernova explosions, the youth of the star-formation episodes may make them
poor candidates in which to make \oVI\ detections.  In contrast, however, the
luminous blue compact galaxy \har\ still emits \oVI\ \citep{Grimes2007}, despite
being under-abundant in nebular metals for its luminosity \citep{Ostlin2015}.
Thus a lower metallicity galaxy may not be a bad candidate per se, provided the
current SFR is high and star formation has been ongoing for some Myr. 

Arguing that both SFR and sSFR should be maximized, we searched the \emph{Sloan
Digital Sky Survey} (SDSS) spectroscopic Data Release 12 \citep{Ahn2014} for
galaxies that have:
\begin{enumerate}
\item{\halpha\ equivalent width (\ewha) above 50~\AA.  This guarantees current
active star formation.}
\item{Redshift in the range 0.23 to 0.29.  This shifts the \oVI\ doublet into
the wavelength region where it can be efficiently observed with HST (this is
described in detail in Section~\ref{sect:method}).}
\item{Less than 0.1 magnitudes of Milky Way extinction in the $u-$band.  This
minimizes foreground extinction that would absorb at far UV wavelengths, and
maximizes the efficiency of the observations. }
\end{enumerate}

\begin{figure}[t!]
\includegraphics[angle=00,width=9cm]{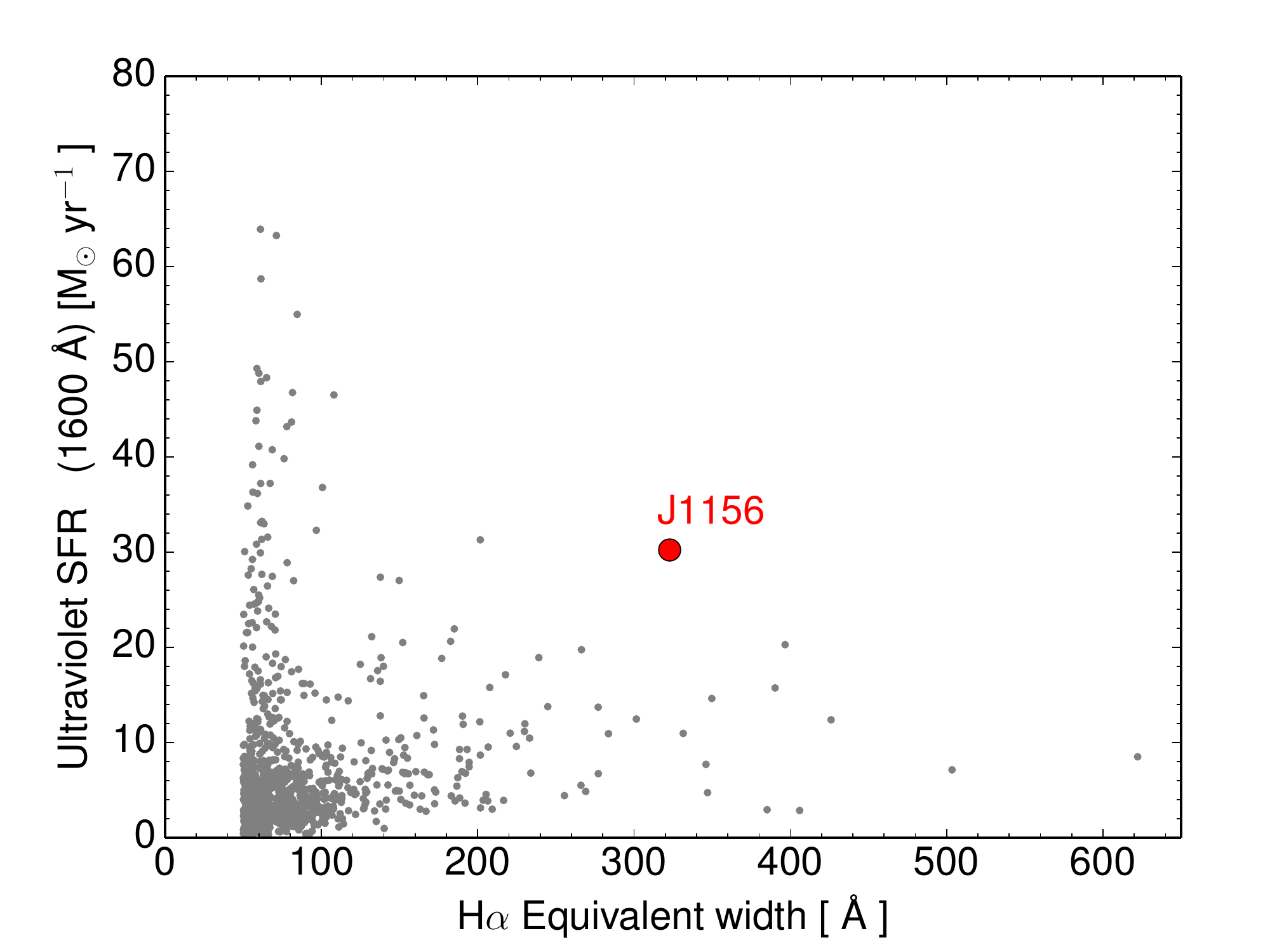}
\caption{The UV SFR vs \halpha\ EW of our SDSS+GALEX catalogues.  The red spot
shows J1156.}\label{fig:select}
\end{figure}

We cross-correlated this first selection with UV photometric catalogues from
Data Release 6 of the \emph{Galaxy Evolution Explorer} (GALEX) archives.  We
computed FUV luminosities using the SDSS redshifts, and converted the result
into star-formation rates \citep[SFR,][]{Kennicutt1998}.  This gave the
selection diagram shown in Figure~\ref{fig:select}.  In this diagram it is clear
that few galaxies have both high SFR and \ewha, and the upper right portion of
the diagram is under-populated: for example, only two objects have both \ewha\
above 200~\AA\ and SFR exceeding 30~\msunyr.  We selected
SDSS\,J115630.63+500822.1, which is shown in red in Figure~\ref{fig:select}. 

\subsection{Basic Properties of J1156}

SDSS\,J115630.63+500822.1 (hereafter J1156) lies at a distance of 1200~Mpc
$(z=0.235)$ and was selected to maximize the chances of detecting \oVI\
emission.  It has a star formation rate $\sim 40$ times higher than the Milky
Way (40~\msunyr), and an optical spectrum that is rich with strong emission
lines suggestive of active star formation.  

\begin{figure}[t!]
\center 
\includegraphics[angle=00,width=60mm]{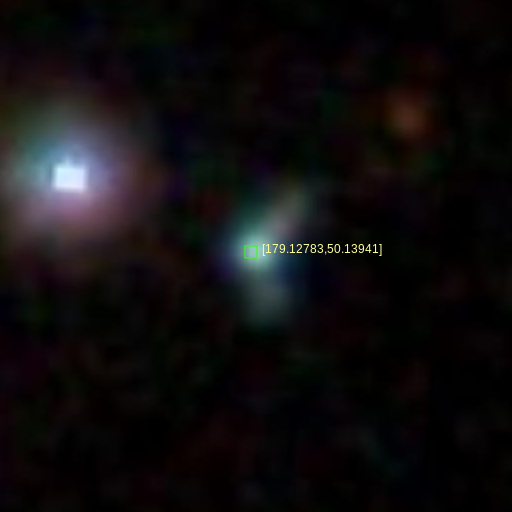}
\includegraphics[angle=00,width=80mm]{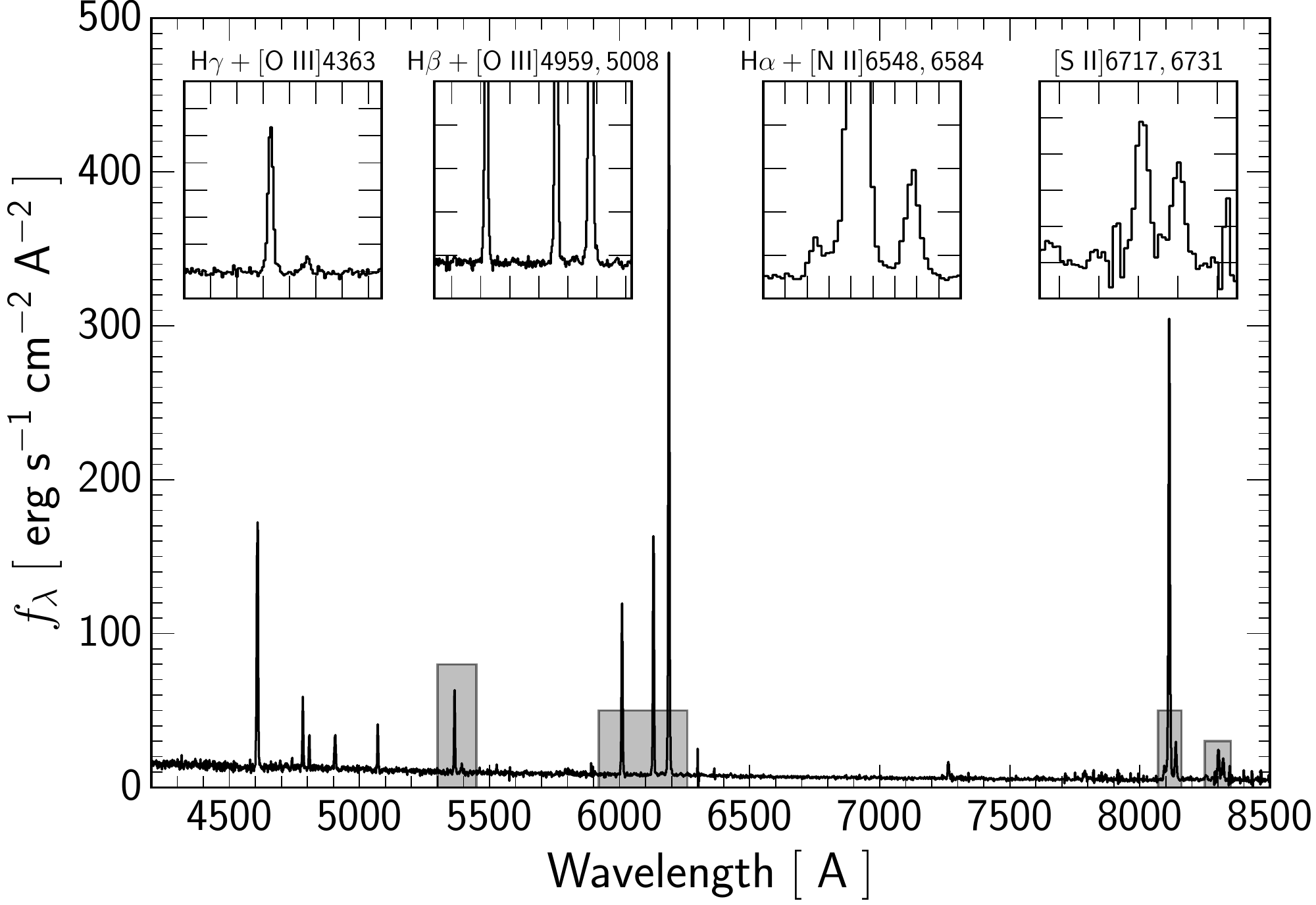}
\caption{SDSS Data for J1156.  \emph{Upper} shows the color composite image, and
\emph{lower} the optical spectrum in the observer frame.  The galaxy is very
compact at optical wavelengths, and the spectrum is characterized by a blue
continuum with very strong, narrow optical emission lines.  Zoomed regions are
illustrated by gray shaded boxes, and show some of the emission lines we use in
this paper.}\label{fig:sdss}
\end{figure}

The SDSS image and spectrum of J1156 are shown in Figure~\ref{fig:sdss}.  We
extract the UV and optical magnitudes from GALEX and SDSS and subtract the
contributions of the strong optical emission lines of [\oII]~$\lambda3727$~\AA,
\hbeta,  [\oIII]~$\lambda\lambda4959,5007$~\AA\ and \halpha\ based upon the
equivalent widths measured the SDSS.  We then proceed to fit the spectral energy
distribution (SED) with with \texttt{Hyper-z} \citep{Bolzonella2000} using
\citet{Bruzual2003} templates with metallicity $Z=0.004$ (closest to the nebular
metallicity) and a Salpeter initial mass function (IMF).  Derived properties are
listed in Table~\ref{tab:props}. 

The FUV absolute magnitude, $M_\mathrm{FUV}$, is --21.2 (AB system), which is
close to the characteristic luminosity, $L^\star$, of Lyman Break Galaxies at
$z=3$ \citep{Reddy2009}.  The UV-to-optical integrated SED is flat in the AB
magnitude system (SDSS $z-$band magnitude is within 0.2 mags of the GALEX FUV
magnitude), indicating a very blue, young stellar population with little
extinction.  The galaxy has a UV SFR of $\approx 30$~\msunyr\ (\halpha\ SFR
$\approx 25$\msunyr, and neither value is corrected for extinction) and a
stellar mass of $\sim 1.5 \times 10^9$~\msun, giving a specific SFR (sSFR) of
$2\times 10^{-8}$~yr$^{-1}$.  Alternatively, J1156 would have built its current
stellar mass in just 50~Myr at the current SFR.  

From the SDSS spectrum we measure the fluxes of a number of commonly used
nebular emission lines.  We compute the \halpha/\hbeta\ ratio to be 3.3, which
is somewhat higher than the value of 2.9 expected for unreddened gas at $10^4$~K
\citep{Osterbrock1989}, and indicates there is little internal extinction of the
nebular gas.  We further compute the oxygen abundance using both the electron
temperature $(T_\mathrm{e})$ sensitive method from the auroral
[\oIII]~$\lambda4363$~\AA\ line and strong-line calibrations, and compute the
electron density $(n_\mathrm{e})$ from the relative strengths of the components
of the [\sII]~$\lambda\lambda6717,6731$~\AA\ doublet using the \texttt{TEMDEN}
task in the \texttt{STSDAS.ANALYSIS.NEBULAR} package of \texttt{Pyraf}.  The
[\oII]~$\lambda\lambda3726,3729$~\AA\ doublet, which also traces $n_\mathrm{e}$,
unfortunately cannot be deblended at the resolution of the SDSS spectrum.  From
the product of $T_\mathrm{e}$ and $n_\mathrm{e}$ we measure the pressure, $P$.
All of these properties are listed in Table~\ref{tab:props}. 

\begin{deluxetable*}{rcccl}[t] \tabletypesize{\scriptsize}
\tablecaption{Derived Properties of J1156\label{tab:props}}
\tablewidth{400pt}
\tablehead{ \colhead{Property}     & \colhead{Derived} & \colhead{$16^{th}$\%-ile$^a$} & \colhead{$84^{th}$\%-ile$^a$} & \colhead{Unit} \\ 
\colhead{(1)}      & \colhead{(2)}      & \colhead{(3)} & \colhead{(4)} & \colhead{(5)} } 
\startdata SFR (UV)$^{b}$          & 18.2  & 18.0     & 18.5  & \msunyr\      \\
SFR (\halpha)$^{b}$                & 26.2  & 25.5     & 27.0  & \msunyr\      \\ 
SFR (\halpha\ corr.)$^{b,d}$       & 39.0  & 33.1     & 44.8  & \msunyr\      \\ 
SFR (SED)$^{c}$                    & 35.3  & 35.2     & 65.4  & \msunyr\ \\ 
Stellar mass (SED)$^{c}$           & 1.5   & 1.4      & 2.5   & $10^9$~\msun  \\ 
SFH (SED)$^{c}$                    & \multicolumn{4}{c}{5 Myr Exponential SFR decline in 65\% of cases} \\ 
Stellar age (SED)$^{c}$            & 11.5  & 10.0     & 39.0  & Myr           \\ 
\ebv\ (SED fit)$^{c,d}$            & 0.14  & 0.13     & 0.15  & Magnitudes    \\ 
\ebv\ (\halpha/\hbeta)$^{d}$       & 0.13  & 0.08     & 0.18  & Magnitudes \\ 
Metallicity (O3N2)$^{e}$           & 8.34  & 8.05     & 8.62  & 12+log(O/H)   \\ 
Metallicity ($T_\mathrm{e}$)$^{e}$ & 7.87  & 7.70     & 8.05  & 12+log(O/H)   \\ 
$T_\mathrm{e}$ ([\oII])$^{e}$      & 1.34  & 1.19     & 1.50  & $10^4$~K      \\ 
$T_\mathrm{e}$ ([\oIII])$^{e}$     & 1.48  & 1.27     & 1.72  & $10^4$~K      \\ 
Electron density [\sII]$^{f}$      & 4.65  & \nodata  & 44.2  & cm$^{-3}$     \\ 
Pressure $(p/k_\mathrm{B})$        & 6.88  & \nodata  & 65.4  & $10^4$~K~cm$^{-3}$ 
\enddata
\tablecomments{Col. 2 represents the derived value, while Cols 3. and 4. present
the limits of the 16th and 84th percentiles.  References and clarifications:
\tablenotetext{a}{Errors on quantities derived direct from photometry or line
ratios are estimated by standard error propagation; errors on $T_\mathrm{e}$ and
SED-fit quantities are derived by Monte Carlo simulations. }
\tablenotetext{b}{SFR \citep{Kennicutt1998} assuming a Salpeter (1955) stellar
initial mass function (IMF) with lower (higher) mass cutoff of 0.1 (100) \msun,
and solar metallicity.}
\tablenotetext{c}{ Properties derived from spectral energy distribution fitting
are calculated using global photometry derived within six Petrosian radii,
calculated in the far UV.  We used the \texttt{Hyper-z} SED fitting code
\citep{Bolzonella2000} using \citet{Bruzual2003} templates  with metallicity
Z=0.004 and the same IMF as above.}
\tablenotetext{d}{Dust corrections use the \citet{Calzetti2000} prescription.}
\tablenotetext{e}{Metallicities are derived using the \citet{Yin2007}
calibration of the O3N2 method, and the algorithm presented in the same paper
for the $T_\mathrm{e}$ method.}
\tablenotetext{f}{Electron density is calculated from the ratio of
$\lambda=6717/6731$~\AA\ components of the [\sII] doublet, using the
\texttt{TEMDEN} task in the \texttt{STSDAS.ANALYSIS.NEBULAR} package of
\texttt{Pyraf}.  The observed ratio is slightly higher than the low-density
plateau \citep{Osterbrock1989}, so we adopt the lowest density quoted. } }
\end{deluxetable*}

\section{Observations with the Hubble Space Telescope}\label{sect:obs}

\subsection{Isolating ultraviolet emission lines in imaging mode}\label{sect:method}

\begin{figure}[t!]
\includegraphics[angle=00,width=9cm]{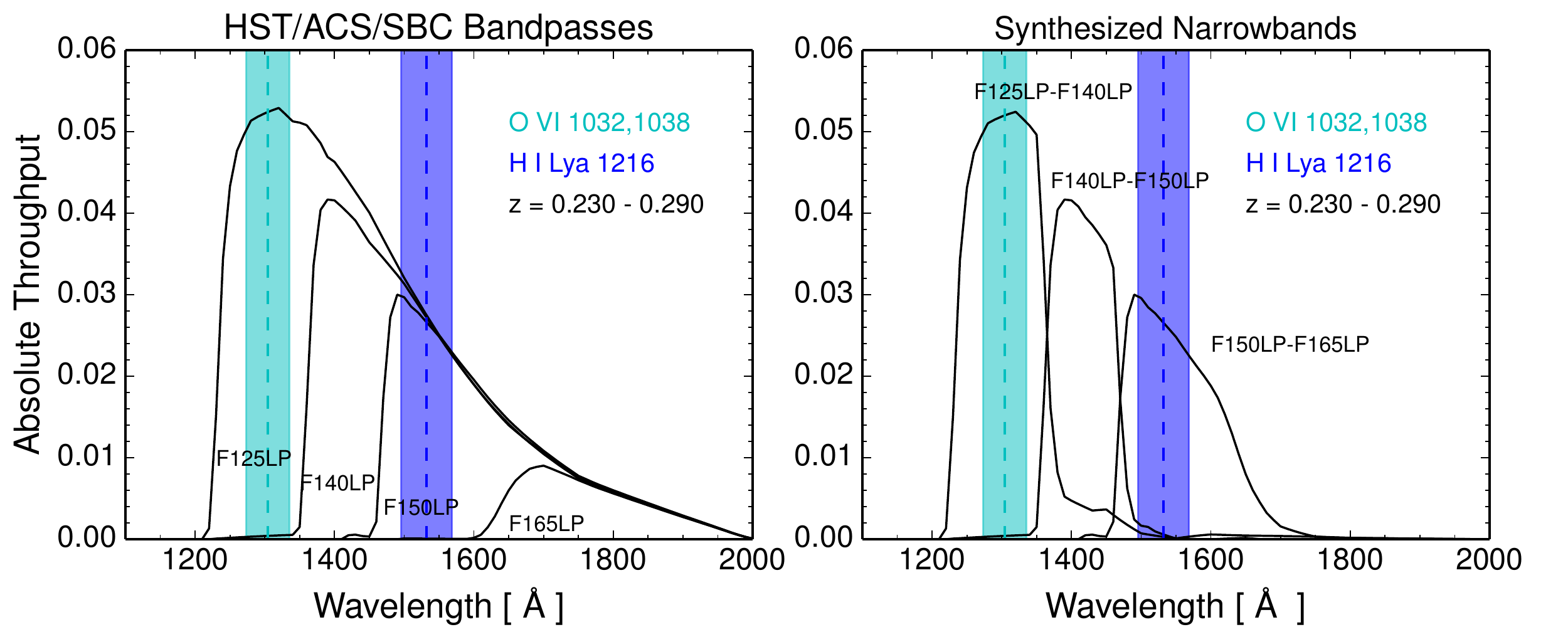}
\caption{The isolation of UV emission lines with ACS/SBC. \emph{Left}: total
throughput of the ACS/SBC filters: F125LP, F140LP, F150LP, and F165LP.  The LP
filters all share the same red wing but have different blue cut-on wavelengths,
so subtraction of adjacent pairs reveals the well-defined narrow bandpasses
(\emph{right}) that we use to isolate \oVI.  At $z = 0.23-0.29$, F140LP
subtracts the continuum from \oVI, which falls only in F125LP.
}\label{fig:filters}
\end{figure}

Isolating line emission requires a narrowband filter, but no conventional
narrowbands exist on the ACS/SBC.  Instead we synthesize them using adjacent
pairs of long-pass filters, as shown in Figure~\ref{fig:filters}.  The long-pass
filters are all described by a sharp, almost step-like, cut-on at the blue side,
and each has an extended red wing that results from the declining quantum
efficiency of the ‘solar blind’ MAMA detector.  The subtraction of adjacent
filter pairs therefore produces well-defined ``pseudo-narrowband'' filters
\citep{Hayes2009}, as illustrated in Figure~\ref{fig:filters}.  The left panel
shows the bandpass profiles, and the right panel shows the profiles that result
from pair-subtraction.  A further advantage of the pair-subtraction is that,
because each bandpass shares the same red wing, each also has the same red-leak
that is then subtracted.  Therefore, no residual signal can be attributed to
red-leaks, as it would be present in all the images taken with long-pass
filters.

We have used this method for continuum-subtracting \hI\ \lya\
($\lambda_\mathrm{rest}=1215.67$~\AA) images in the \emph{Lyman Alpha Reference
Sample} \citep[LARS,][]{Hayes2014,Ostlin2014} in total we have applied it to
around 50 starburst galaxies selected to redshift \lya\ into the narrowband
windows.  In designing the \oVI\ study we used the same strategy, but instead
applied a redshift constraint to shift the \oVI\ doublet into the F125LP--F140LP
window.  This sets the redshift range of $z = 0.23-0.29$, as mentioned in
Section~\ref{sect:target}, and conveniently also redshifts the \hI\ \lya\ line
into the wavelength region isolated by the F150LP-–F165LP pair
(Figure~\ref{fig:filters}), which we also obtained.

\subsection{HST Observations and Strategy}

\subsubsection{Imaging Data}\label{sect:obs:imgdata}

J1156 was observed in both imaging and spectroscopic modes.  Data can be found
in the \emph{Mikulski Archive for Space Telescopes} (MAST), under GO\,13656.
Table~\ref{tab:obs} summarizes the observations, which were performed with: 
\begin{itemize}
\item{The \emph{Advanced Camera for Surveys} (ACS), using long-pass filters of
the Solar Blind Channel (SBC) in the ultraviolet, and tunable linear ramp
filters of the Wide Field Channel (WFC) in the optical.}
\item{\emph{The Wide Field Camera 3} (WFC3), using UVIS broadband filters in the
optical.}
\item{\emph{The Cosmic Origins Spectrograph} (COS), using the FUV channel and
the G130M grating to obtain medium resolution spectroscopy. }
\end{itemize}

\begin{deluxetable*}{ccccl}[t]
\tabletypesize{\scriptsize}
\tablecaption{HST Observing Log.\label{tab:obs}}
\tablewidth{0pt}
\tablehead{
\colhead{Instrument} & \colhead{Filter/Grating} & \colhead{Central Wavelength
(\AA)} &
\colhead{Exposure Time (s)} & \colhead{Purpose / Comment} \\ 
\colhead{(1)} & \colhead{(2)} & \colhead{(3)} & \colhead{(4)} & \colhead{(5)} 
}
\startdata
ACS/SBC   & F125LP & 1438 & 15,864 & \oVI\ online / 6 of 8 visits    \\
ACS/SBC   & F140LP & 1528 & 17,304 & \oVI\ continuum / 6 of 8 visits \\
ACS/SBC   & F150LP & 1612 & 2,742  & \lya\ online                    \\
ACS/SBC   & F165LP & 1763 & 5,442  & \lya\ continuum                 \\
WFC3/UVIS & F390W  & 3923 & 912    & Restframe $U-$band              \\
WFC3/UVIS & F475W  & 4773 & 700    & Restframe $B-$band              \\
ACS/WFC   & FR601N & 6010 & 1,419  & \hbeta\ narrowband              \\
ACS/WFC   & FR782N & 8111 & 700    & \halpha\ narrowband             \\
WFC3/UVIS & F850LP & 9167 & 700    & Restframe $I-$band              \\
\hline
COS/FUV   & G130M  & 1318 & 2,360  & 1150-—1450~\AA\ spectroscopy 
\enddata
\tablecomments{Imaging filters are ordered by increasing central wavelength; the
COS spectrum is the last entry.   For the F125LP and F140LP frames, only data
from six of the eight visits was used because of noticeable dark current in the
other two.  }
\end{deluxetable*}

Observations of \oVI\ and its adjacent continuum must be particularly deep: for
this program we have dedicated 16 spacecraft orbits to (combined) imaging in
F125LP and F140LP, which were divided into eight visits of two orbits each.  We
worked closely with the contact scientists and scheduling team at the Space
Telescope Science Institute (STScI) to ensure that these visits were executed
only when the SBC had not been used during the previous 24-hour period.  This
was to guarantee that the readout amplifiers had been switched off for a
significant time \footnote{ACS Instrument Science Report ACS 2009-02}, which
allows them to cool.  This reduces the dark current, which is strongly
temperature-dependent, to a level that is below the background.  However this
visit execution requirement was not fulfilled for two of the F125LP+F140LP
visits (02 and 05), resulting in significantly increased background.  We
excluded data that were obtained during these two visits. 

For F125LP+F140LP we executed a 16-point dithering pattern, with two positions
performed in each of the eight visits.  The 16 points dither over a $4\times
4$-pointing grid, stepped at each position by 3 arcsec in detector coordinates.
No orientation constraint was placed on the roll angle of the telescope, and
hence not only are the dithers large, but the roll angle also varies.  This
strategy was designed to minimize residuals from the flatfielding by averaging
over 16 well-separated detector positions.  In reality, only 12 pointings were
used for the final stacked images because two visits were removed.  In each
orbit we executed the F125LP observation during orbital night (`shadow') to
minimize the background due to the geocoronal oxygen emission lines near
1304~\AA.

SBC imaging of \lya\ and the adjacent continuum (see Figure~\ref{fig:filters})
were executed in one orbit for F150LP and two orbits in the less sensitive
F165LP.  WFC3/UVIS imaging in F390W, F475W, and F850LP was accomplished in a
single orbit, using two exposures for each filter and a half-pixel dithering
strategy in x and y detector coordinates.  ACS/WFC ramp filter imaging to obtain
\halpha\ and \hbeta\ narrowband frames were also accomplished in one orbit with
half-pixel dithering.  For WFC3 and ACS CCD imaging we post-flashed each
exposure with the number of electrons needed to reach the recommended background
levels that minimized the charge transfer inefficiency at read-out. 

\subsubsection{Spectroscopic Data}

Spectroscopic data with the COS were taken with the G130M grating, which
provides medium resolution spectroscopy between 1150 and 1450~\AA.  The
effective area of G130M reaches a maximum near 1250~\AA, precisely the
wavelength region to which our narrowband imaging using F125LP--F140LP is tuned.
The grating provides a spectral resolving power (assuming a point source) of
$R\approx 20,000$ at 1250~\AA, or around 15~\kms. This enables us to resolve
interstellar emission and absorption profiles on the galaxy scale, and indeed
all lines are clearly resolved in the spectra.  COS observations were
accomplished in a single orbit using the 1309~\AA\ central wavelength setting.
All four FP-POS settings were used to dither each spectral image in the
dispersion direction and average over fixed-pattern noise, gain sag, and remove
grid wire shadows.  The target acquisition worked flawlessly, placing the
brightest UV point in the center of the Primary Science Aperture to within
0.05~arcsec.

\section{Data Reduction and Processing}\label{sect:reduct}

\subsection{Imaging Data}

\subsubsection{Initial Reduction of the Science and Weight Frames}

We began our data reduction process by obtaining all images from the MAST,
taking data that had been pipeline-processed to the point of flat-field
correction (i.e. the \texttt{\_raw.fits} and \texttt{\_flt.fits} frames). 

CCD channels (WFC3/UVIS and ACS/WFC) have imperfect charge transfer efficiency,
producing streak-shaped artefacts around bright objects (and mostly cosmic ray
hits), particularly if the dominant source of noise is from readout.  While we
did post-flash our observations to minimize this, many of these streaks remain
in the individual \texttt{flt} images.  For the ACS/WFC images, a postiori CTE
correction is performed by the pipeline, but for the WFC3/UVIS frames this is
not the case.  For UVIS images we process the \texttt{flt} images with the
\texttt{wfc3uv\_ctereverse\_parallel} code, developed at STScI.

The next step is to register all the frames onto a common pixel grid and co-add
frames from each filter.  For this we make use of the \texttt{drizzlepac}
software, where we used \texttt{astrodrizzle} to correct for distortions and
register all frames onto a grid of 0.1 arcsec pixels.  Small shifts in the world
coordinate system, that arise from the use of different guide stars in different
visits, were rectified with the \texttt{tweakreg} package, and images were
re-drizzled from scratch to minimize the number of times the pixels were
resampled.  All frames were then matched to the F390W filter, as it is the most
central filter in (logarithmic) wavelength space.  The inverse variance weight
map for each filter was saved in addition to the combined science image. 

We performed an additional sky subtraction in the reduced science frames.  We
computed the sky values in boxes of several square arcsec, which were positioned
around the galaxy ($\sim10$~arcsec away), and avoided obvious features.  We then
approximated the sky by a flat surface that could vary in tilt, described by a
first order polynomial in x and y with no cross-term.  This surface was
subtracted from the science frames, providing residual sky that was an order of
magnitude below the sky noise in the same boxes where the sky was measured.

\subsubsection{Matching the Point Spread Functions}\label{sect:psfmatch}

We estimate the line-only flux in the \oVI\ lines using several methods, that we
describe in Section~\ref{sect:contsub}.  Some of these methods require the use
of longer-wavelength data, obtained from the WFC3/UVIS and ACS/WFC cameras, and
hence rely an precise matching of the point spread functions (PSF) of the
images.  This is particularly important because the PSF of the SBC moves
significant power to radii larger than $\sim 0.5$~arcsec, and produces extended
wings that are not present in data from the optical cameras.

We first construct PSF models for all of the filters used in the study. For the
optical filters we use TinyTim models which are placed into the (un-drizzled)
pipeline-processed \texttt{flt} frames. We then drizzle the frames in the same
way as the actual data frames and extract PSF models from the drizzled output.
This procedure does not work for the FUV data obtained with SBC because the
TinyTim models are not accurate enough (especially not in the extended wings).
For the SBC data we instead build PSF models from stacks of stars observed in a
stellar cluster (see Appendix~\ref{sect:psf} for details). We then proceed to
make a maximum width PSF model by normalizing all models to the same peak flux
and stack them by maximum pixel value. 

We then find convolution kernels that match the PSFs for all of the filters to
the maximum width model. Each kernel is built up from a set of delta functions
and we find the optimum matching kernel by least squares optimization. The
method is similar to the technique described in \citet{Becker2012} and details
will be presented in a forthcoming paper (Melinder et al. in preparation). The
method is very sensitive to noise in the input PSFs and using models is
therefore required to compute the matching kernel. The drizzled and registered
images are then convolved with the optimal matching kernels which result in a
set of images matched to a common PSF.

\subsubsection{Voronoi Tessellation}

As is often the case for imaging of extended sources, the signal-to-noise ratio
(SNR) is high in the central regions but low in the outskirts.  In this study we
are particularly concerned about the lower surface brightness, diffuse regions.
We therefore use adaptive binning tools to aggregate pixels together as a
function of local SNR, until a threshold SNR is met.  Hence in the central
regions of the galaxy bins are small, and the exquisite spatial resolution of
HST is preserved, while in the outskirts we sacrifice some of the spatial
resolution in favor of higher SNR.  The advantage of binning, as opposed to
smoothing, is that every pixel contributes once and only once to a bin, and thus
mathematical treatment of the noise is straightforward. 

We made use of the \texttt{Weighted Voronoi Tessellation} binning algorithm by
\citet{Diehl2006}, which is a generalization of the \citet{Cappellari2003}
Voronoi binning algorithm.  We adopted the F125LP image as a reference image
because it contains both UV continuum and \oVI, and binned together pixels until
either a minimum SNR of 10 was obtained, or bins reached a maximum size of 100
pixels (1 square arcsecond).  With the binning pattern computed for the F125LP
frame, we applied the same pattern to images taken in all the other filters,
again storing the science image and the inverse variance weight map.  All
photometric quantities reported in this article are derived from the tessellated
frames.

\subsubsection{Subtracting the Continuum from \oVI}\label{sect:contsub}

The technique described in Section~\ref{sect:method} isolates a narrow
bandpasses in the UV, but we remain concerned about several issues that enter
the estimation of the continuum in F125LP filter.  In previous studies where we
have targeted \lya, we have explored and discussed different methods of
continuum-subtracting SBC imaging data, using various assumptions and models
\citep{Hayes2005,Hayes2009}.  Specifically the filter that samples \oVI\ is
broad compared to conventional narrowband filters, and an extrapolation is
required between the continuum-only F140LP filter and F125LP.  This
extrapolation must be done in the UV, where the continuum shape may vary rapidly
as a function of stellar age and dust reddening, and therefore spatially.  In
regions where the stellar continuum is bright, small errors in the estimated
F125LP continuum level may result in significant residual flux (either positive
or negative) when the continuum is subtracted.  In continuum-bright regions the
UV slope or stellar properties may be the best constrained but,
counterintuitively, the line flux estimates are in fact the least precise.

The line-flux, $F_\mathrm{line}$ is calculated from the following expression:
\begin{equation}
F_\mathrm{line} = (f_\mathrm{125} - \zeta_\mathrm{UV} \cdot f_\mathrm{140})
\cdot W_\mathrm{125}
\end{equation}
where $f_\mathrm{125}$ and $f_\mathrm{140}$ are the $f_\lambda$ flux densities
measured in the F125LP and F140LP filters, respectively, and $W_\mathrm{125}$ is
the bandwidth of the F125LP filter.  The critical quantity is
$\zeta_\mathrm{UV}$, which is the factor that scales the F140LP flux to estimate
the contribution of continuum processes to the F125LP flux.  $\zeta_\mathrm{UV}$
is similar to the \emph{Continuum Throughput Normalization} (CTN) factor defined
in \citet{Hayes2005,Hayes2009}, but is here quoted in flux units without folding
in the filter throughput.  $\zeta_\mathrm{UV}$ is a single number that captures
a lot of physical quantities; to first order it reflects the shape of the UV
continuum.  This slope is usually parameterized using the functional form of a
power-law: $f_\lambda \propto \lambda^\beta$, where $f_\lambda$ is the continuum
flux density per wavelength, which casts $\beta$ as the (conventionally defined,
logarithmic) UV spectral slope.  This is governed by stellar age, metallicity,
and reddening.  Adding complexity, $\zeta_\mathrm{UV}$ may also encode
information about discrete features that may fall within the on-line bandpass. 

\begin{figure*}[t!]
\includegraphics[angle=00,width=18cm]{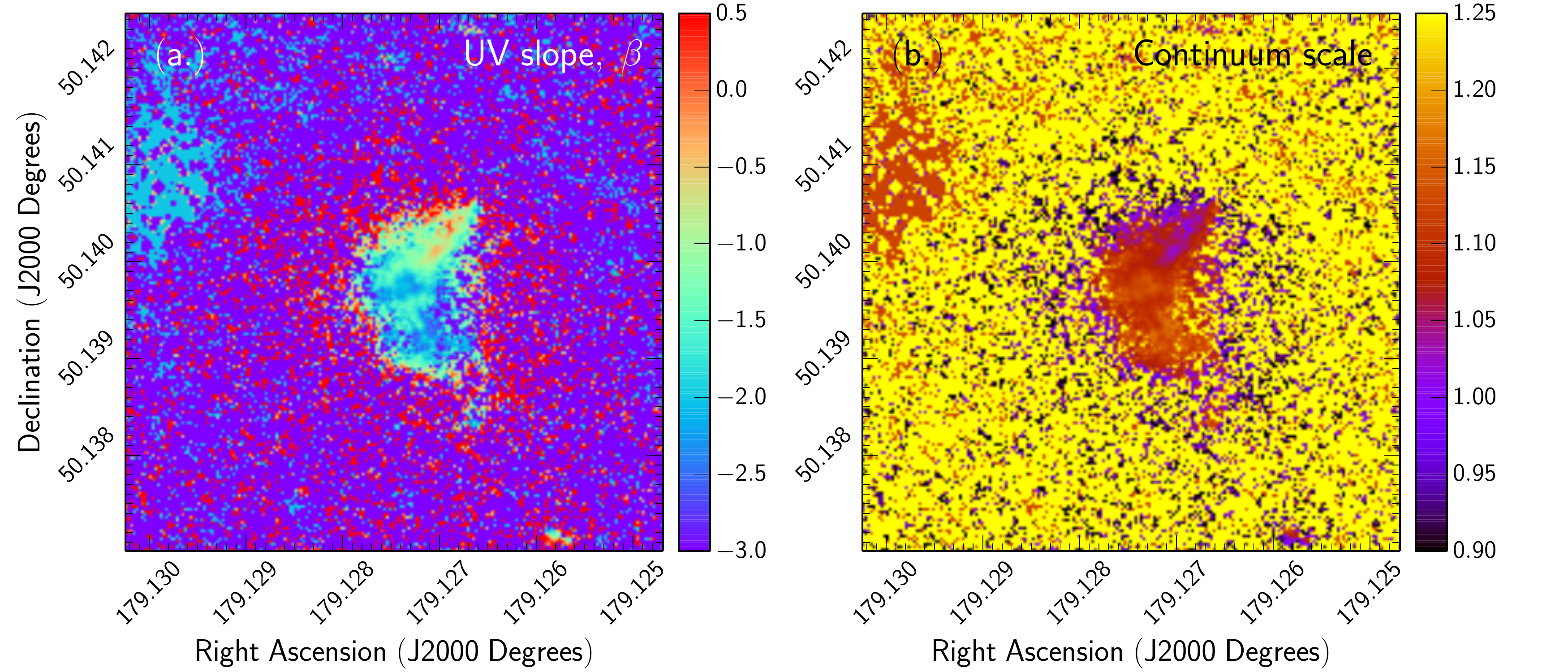}
\caption{The UV Color of the Stellar Population. \emph{Left} shows the UV
continuum slope $\beta$, which can be seen to range from as red as 0.5 to as
blue as $-3$.  \emph{Right} shows the corresponding factor by which the offline
filter (F140LP) would need to be scaled (in $f_\lambda$ units) to produce the continuum
flux in F125LP for the same $\beta$ slope, assuming no line emission or absorption in
the blue remainder of the bandpass.  }\label{fig:beta}
\end{figure*}

We begin by evaluating the UV slope empirically as a function of position, which
we show in Figure~\ref{fig:beta}.  We take the images in ACS/SBC/F125LP, F140LP,
F150LP, F165LP and WFC3/UVIS/F390W, which span the spectral region between
1160~\AA\ and 3100~\AA\ in the restframe (pivot wavelengths for each filter).
Including the weight maps output by \texttt{astrodrizzle}, we perform a weighted
least-squares fit in every pixel, giving the `$\beta$-map' shown in left panel
of Figure~\ref{fig:beta}.  This shows that in the central nuclear condensation,
the UV slope has a logarithmic gradient of $\beta\approx -1.5$ to $-2$, which is
consistent with a young, mostly unreddened starburst, although note that this
measurement covers a longer wavelength baseline than normally defined.  Some
regions show $\beta$ at least as blue as $-2.5$ at high significance, which is
not easy to reproduce with stellar models if all stars are cospatial.  For
example in some strong starbursts (e.g. Tololo\,1247--232; \citealt{Buat2002};
Puschnig et al in preparation), the UV continuum shortwards of \lya\ is too blue
to be describable with spectral models, a problem that may be exacerbated when
smaller regions are resolved (e.g. at HST resolution).  In fainter regions --
particularly the northern extended structure -- the UV color is much redder and
shows $\beta$ around $-0.5$ to 0.5.  This is also clearly visible in
Figure~\ref{fig:oviimg}, where the color composite image shows this arm to be
much redder than the rest of the galaxy.  A ring of pixels showing a UV slope
around $\beta=0$ (flat in $f_\lambda$) encompasses the galaxy; most likely this
is because the starburst is hosted in an older stellar population.

In this paper we test three methods of estimating $\zeta_\mathrm{UV}$, which
each has advantages and disadvantages. In order of increasing complexity, these
are:

\paragraph{A single power-law \textsc{(single-PL)}} This is the simplest method,
in which we adopt a single value for $\zeta_\mathrm{UV}$ across the whole image.
We adopt a fiducial value of $\beta = -1.9$, it being the total value computed
in a global aperture.  This corresponds to $\zeta_\mathrm{UV}=1.12$.  Adopting a
single power law enables us to rapidly explore the limitations due to systematic
errors, which we do in  Section~\ref{sect:res:sign} using a range of observed
$\beta$ values.

\paragraph{A spatially varying power-law \textsc{(pixel-PL)}} As shown in the
left panel of Figure~\ref{fig:beta} the continuum slope is not constant over the
whole galaxy, and varies between $\beta=-3.4$ and $\beta=0$.  Assuming the
power-law is a complete description of the continuum shape, this corresponds to
a range in scale factor, $\zeta_\mathrm{UV}$, of 1.23 to 1.0.  In this method we
use the map of $\zeta_\mathrm{UV}$ to subtract a continuum level that varies as
a function of position.  A key point to note from these Figures is that when the
UV slope is \emph{blue} $\zeta_\mathrm{UV}$ is \emph{high}; this results in
\emph{more} continuum being subtracted and \emph{less} inferred line emission.
Redder slopes would result in more inferred line emission. 

\paragraph{A model SED in every pixel \textsc{(pixel-SED)}} In reality the UV
continuum is not shaped by a single equation but by the properties of stars,
gas, and dust;  we are therefore concerned that even the \textsc{pixel-PL}
method is insufficient.  To estimate the effects of this we implement a more
sophisticated method of extrapolating the continuum into the blue end of the
F125LP bandpass using the spectral models.  This is the same method we use to
subtract the UV continuum from \hI\ \lya\ observations in LARS; see
\citet{Ostlin2014} for the latest description.  In short we perform a stellar
SED fit in every pixel, using the \emph{Starburst99} synthetic spectral models
\citep{Leitherer1999,Leitherer2014}.  We adopt a constant stellar metallicity
that is closest to the measured nebular value ($Z=0.004$).  Using the five bands
of continuum imaging we fit two components of stars, after iteratively
subtracting the nebular gas spectrum using the narrowband observations of
\halpha\ and \hbeta.  In each pixel we extrapolate the best-fitting spectrum,
convolve it with the F125LP bandpass, and subtract the result from the
observation.  We note that this method does not perform as well as it does in
the LARS project, because the age determination requires the filters to closely
bracket the Balmer and 4000~\AA\ spectral breaks.  At this redshift the F390W
filter is well positioned on the blue side, but the F475W filter is centred at
3900~\AA\ in the restframe, and does not well-sample the red side.

\medskip 
Each method has its own advantages and disadvantages.  The \textsc{single-PL}
method allows for very fast tests to be performed, and to estimate the absolute
limits of systematic uncertainty.  The \textsc{pixel-PL} method enables us to
account for spatially varying UV colors but exploring the extreme colors is less
certain.  Neither of the first two methods actually includes any physics, and
both assume the continuum behaves as a power-law.  The measured values of
$\zeta_\mathrm{UV}$ vary by only $\approx 20$~\% (minimum to maximum) across the
full image, and the variation is significantly smaller where the UV is well
exposed.  The \textsc{pixel-SED} method allows for realistic shapes of the
continuum (e.g. stellar continua and dust reddening do not necessarily behave
like power-laws) and may even account for discrete features if the age is
constrained precisely enough.  However spectral models for O stars are uncertain
in the far UV and even more so at metallicities below that of the Magellanic
clouds; J1156 has a metallicity around 1/10 the solar value \citep{Asplund2004}.
For the \textsc{pixel-SED} to perform accurately, the shape of assumed
dust-reddening curve must also accurately reflect that of the galaxy.

We are cognizant of the fact that we are trying to measure a line that has a net
equivalent width of a few tenths of an \AA.  The bandpass we use is rather broad
at $\sim 100$~\AA, and where the UV continuum is bright and the local EW is low,
we do not have the contrast to reliably measure the \oVI\ feature whatever
technique we use.  However where the continuum is faint and the emission
extended, the local EW can become almost arbitrarily high, and if there is
little continuum to subtract, then this error is minor and absorption from the
continuum is negligible.  In such regions the flux will be quite precise, but
the local EW will still remain poorly measured.

In this article we present continuum subtractions based upon all three methods.
However all of our photometry is based upon the \textsc{single-PL} method which
allows us to estimate the errors per pixel and rapidly quantify both systematic
and statistical errors.  

We estimate statistical errors with Monte Carlo simulations.  We use the
\texttt{wht} image from \texttt{astrodrizzle} and convert it to a standard
deviation per pixel.  We account for the fact that the drizzling process
correlates the noise, and apply the standard average correction
\citep{Casertano2000,Fruchter2002}.  We re-generate the Voronoi cells in the
F125LP and F140LP images, and perform the same continuum subtraction using the
fiducial UV slope of $\beta = -1.9$.  We produce 1000 realizations of the \oVI\
image, and for each tessellated cell compute the upper and lower limits that
encompass the central 68th percentile ranges of the distribution. 

\subsection{Spectroscopic Data}\label{sect:obs:spec}

Target acquisition with the COS was done in \texttt{ACQ/IMAGE} mode.  We first
checked that the peak-up process had well centered the galaxy within the 
Primary Science Aperture (PSA). We examined the header keywords
\texttt{ACQPREFX} and \texttt{ACQPREFY} of the \texttt{\_rawacq} frame, which
records the pixel upon which the PSA is centered after the telescope slew.  We
determine the pixel centering to be just 2 pixels (0.05~arcsec, in a 2.5~arcsec
aperture) from the intensity centroid of the brightest condensation in the near
ultraviolet acquisition frame. 

With the object centered in the aperture, we processed the data using the
default settings of the \texttt{CALCOS} pipeline, stopping after spectral
extraction and the generation of four \texttt{\_x1d} files (one for each
\texttt{FP-POS} setting).  We resampled each 1D spectrum onto a common
wavelength grid and combined them with the \texttt{splice} task in the
\texttt{stsdas.hst\_calib.ctools} package of \texttt{Pyraf}.  In this step,
unlike the complete run of \texttt{CALCOS}, we more conservatively rejected all
spectral pixels that had a non-zero data quality flag. We checked that the
geocoronal \lya\ and \oI\ line have wavelength centroids that correspond to the
vacuum wavelength of the transition; for \lya\ they agree to within 0.01~\AA\
(2.5~\kms).

\section{Description of the Results}\label{sect:res}

\begin{figure*}[t!]
\includegraphics[angle=00,width=18cm]{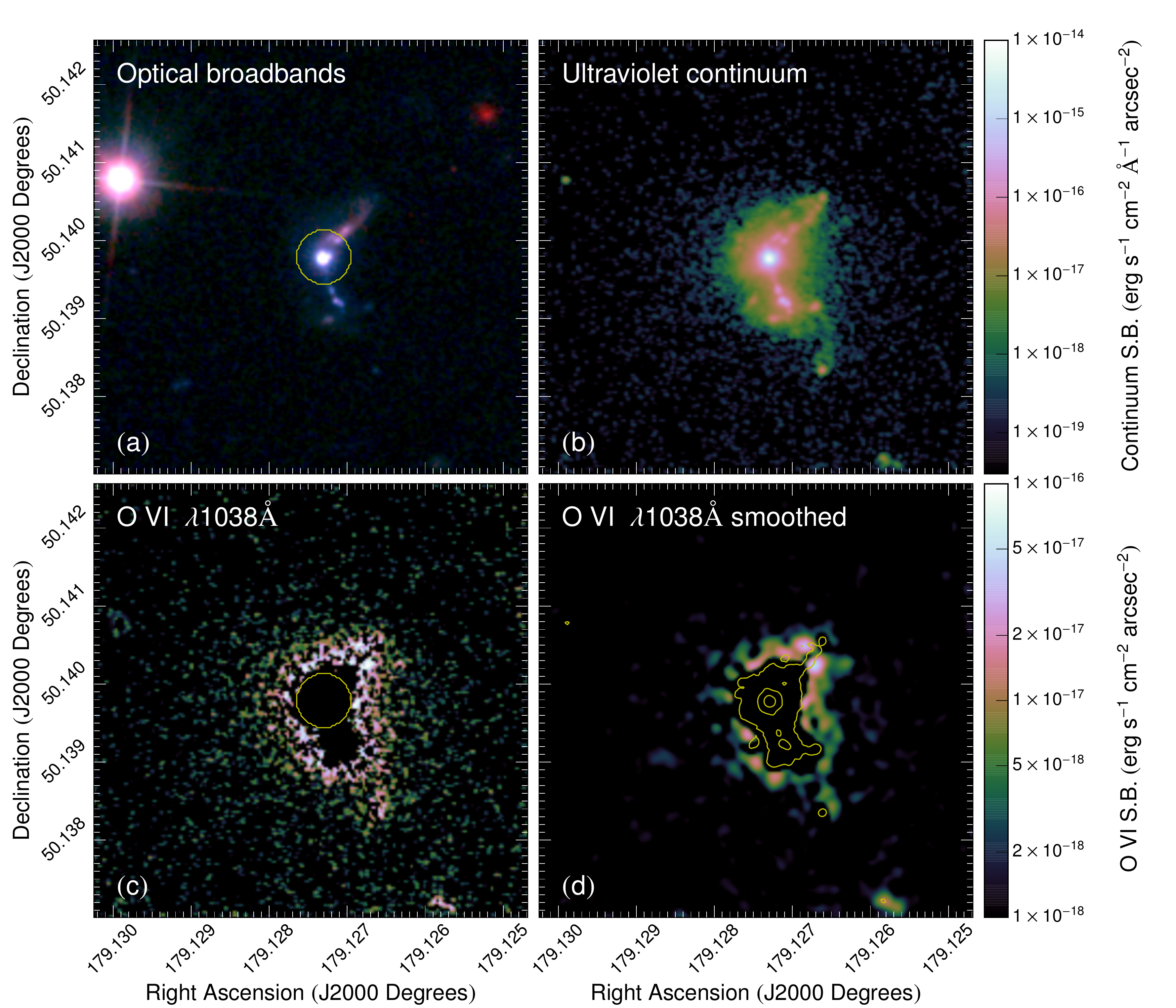}
\caption{Imaging of J1156. \emph{Upper left}: A color-composite image: the blue
channel encodes the F390W filter (restframe 3150~\AA), green shows the F475W
filter (restframe 3900~\AA), and red shows F850LP (restframe 7200~\AA).  The
yellow circle shows the position and size of the spectroscopic aperture.
\emph{Upper right} An ultraviolet continuum image (restframe 1250~\AA).
\emph{Lower} panels show the emitted line flux that we interpret to be the
\oVI~$\lambda 1038$~\AA\ emission line.  Images use a logarithmic scale.  The
\emph{left} panel is un-smoothed, and the \emph{right} is smoothed with a
Gaussian kernel with $\sigma=3$ pixels. The yellow contours show UV continuum
radiation at levels of {$10^{-17}$, $10^{-16}$,
$10^{-15}$}\ergseccmarcsec.}\label{fig:oviimg}
\end{figure*}

The \emph{upper left} panel of Figure~\ref{fig:oviimg} shows a color composite
of the galaxy, taken in HST optical filters.  It has an extremely compact core,
just 1~kpc across in the FUV image (\emph{upper right}) that hosts almost all
the star-formation.  Fitting an exponential profile to the UV continuum gives a
scale radius of just $\approx 0.75$~kpc, in which the SFR is of the order of
30~\msunyr.  The galaxy also shows extended streams indicating that its origin
may lie in the merger of two galaxies.  

\begin{figure*}[t!]
\includegraphics[angle=00,width=18cm]{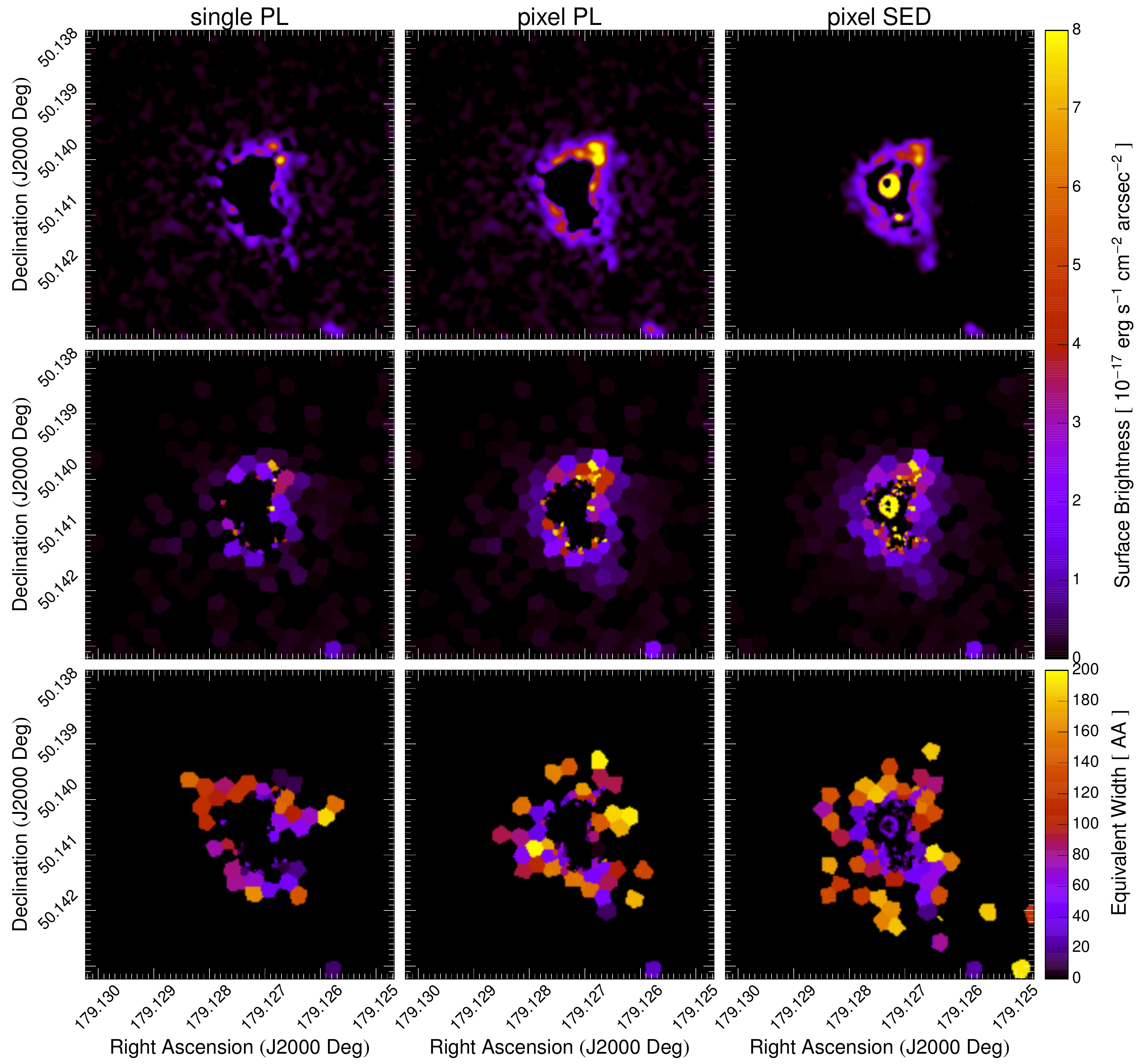}
\caption{A comparison of the different methods used to subtract the continuum.
The \emph{upper} row shows the \oVI\ surface brightness estimated in the
unbinned data that have then been smoothed with a 3 pixel Gaussian kernel; the
\emph{central} row shows the corresponding surface brightness in the Voronoi
tessellated data.  The \emph{lower} row shows the \oVI\ equivalent width
estimated from the Voronoi tessellated data.  Colorbars for each row of Figures
are shown to the right.  The columns show compare the different methods:
\emph{left} shows the {\sc single-PL} method, \emph{centre} shows the {\sc
pixel-PL} method, and \emph{right} shows the {\sc pixel-SED} method.  See text
for details.}\label{fig:methcomp}
\end{figure*}

\subsection{\oVI\ Absorption and Emission }\label{sect:res:obs}

Continuum-subtracted F125LP images are shown in the \emph{lower} panels of
Figure~\ref{fig:oviimg}.  Figure~\ref{fig:methcomp} shows a set of the same
continuum-subtracted images, in which we compare the three methods described in
Section~\ref{sect:contsub}: \textsc{single-PL}, \textsc{pixel-PL}, and
\textsc{pixel-SED}.  These images also compare the morphology of the
continuum-subtracted data when performed on unbinned pixel images and then
smoothed, and when Voronoi tessellation is applied.  The \emph{lower} panels
also include maps of the equivalent width (EW) in the tessellated data only. 

The central regions of the galaxy show negative flux. This is expected for
observations of strong electronic transitions superposed against bright
background continuum sources (as with the absorption line studies discussed in
the Introduction), as \hI\ (\lyb) and \oVI\ absorb radiation from the continuum.
However, a ring of positive flux is clearly visible surrounding the UV-bright
regions of the galaxy.  The morphology is not symmetric, and is brighter towards
the northwest.  In Figure~\ref{fig:methcomp}, which compares continuum
subtractions performed by various methods, we find a close agreement in the flux
produced by the {\sc single-PL} and {\sc pixel-PL} methods.  The {\sc pixel-SED}
method, however, produces a bright region emission from the central
condensation.  The ring-like halo emission, however, shows a similar surface
brightness in all methods.  The bright central emission is absolutely not
supported by the COS spectrum, and this spurious emission arises because the
stellar models at this wavelength cannot fully capture the shape of the
continuum across the bandpass.  We do not consider continuum subtractions made
using this method further in this paper.  The {\sc pixel-PL} method produces a
narrow ring of higher surface brightness than {\sc single-PL}, but note also in
Figure~\ref{fig:methcomp} that the absorbing region is not as wide in the {\sc
pixel-PL} subtraction.  The equivalent width in such imaging data is
particularly hard to measure, as both the emission line and the continuum are
faint at radii where the \oVI\ emission dominates over \lyb\ and other
absorption.  Minor differences in scale factor produce large relative
differences in the continuum-subtracted flux, and EWs measured in individual
pixel bins are quite uncertain.  The main point to read from these EW maps is
that the lower values (purple in the color scaling) are found closer to the
galaxy, where the continuum is brighter and the high EWs  (yellow, red) are
found at larger radii.  These pixels at large radii, however, are also the most
uncertain.

\begin{figure}[t!]
\includegraphics[angle=00,width=9cm]{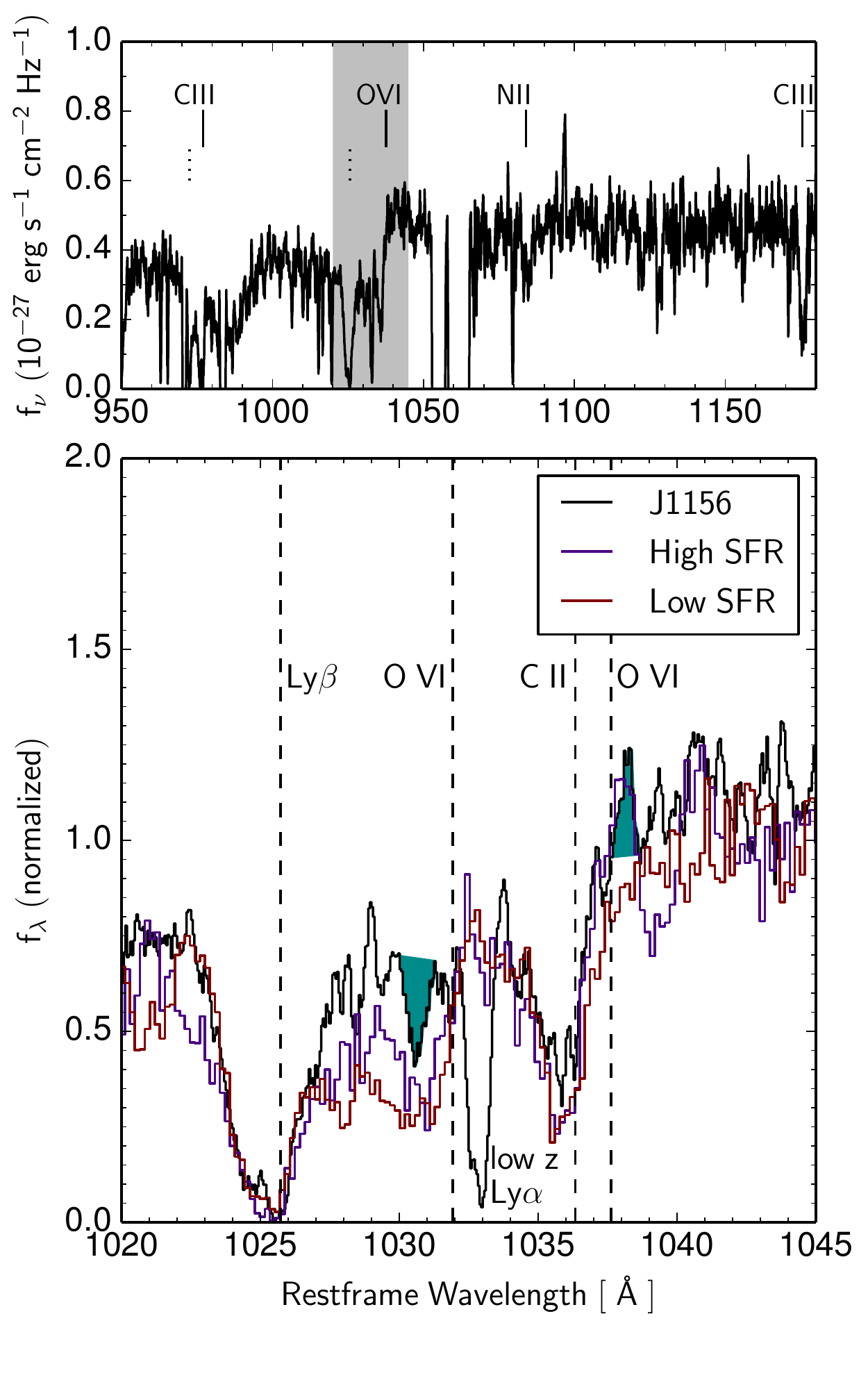}
\caption{The ultraviolet spectrum around \lyb\ and \oVI. The \emph{upper} panel
shows the full spectrum, shifted into the restframe.  Strong interstellar metal
absorption features are labeled; Lyman lines are shown by vertical dotted lines.
The \emph{lower} panel shows the \oVI\ and \lyb\ region (shaded above) of J1156
in black.  Filled patches show the regions identified as the $\lambda 1032$~\AA\
absorption line (blueshifted) and $\lambda 1038$~\AA\ emission line.  The
average spectra of strongly and weakly star-forming galaxies are shown in purple
and red, respectively \citep{Heckman2011,Henry2015}. The spectrum of J1156 and
strongly star-forming subsample are remarkably similar and both show a clear
\oVI\ emission line at $\lambda 1038$~\AA. }\label{fig:ovispec}
\end{figure}

The most likely interpretation of this ring of emission is that the emission
comes from the \oVI\ doublet because, very importantly, there are no other
strong recombination or collisional lines in the bandpass.  The interpretation
of \oVI\ emission is supported by a spectrum, obtained in the center, which we
show in Figure~\ref{fig:ovispec}.  The spectrograph has a 2.5-arcsecond diameter
circular aperture, which was centered upon the UV dominant star cluster, as
illustrated in Figure~\ref{fig:oviimg}.  The spectrum shows saturated absorption
in \lyb, and no emission.  \oVI\ is also primarily in absorption; the $\lambda
1032$~\AA\ part of the doublet is clearly outflowing, and we measure a blueshift
of 380~\kms.  The $\lambda 1038$~\AA\ feature is blended with \cII~$\lambda
1036$~\AA\ absorption, and combined these absorption features confirm the
central negative flux seen in Figure~\ref{fig:oviimg}.  

The spectrum shows a weak emission line at restframe $\lambda 1038$~\AA, with an
integrated flux of $(1.5 \pm 0.6) \times 10^{-16}$~\ergseccm.  The flux
calibration applied by the COS pipeline assumes a point source in the centre of
the aperture.  This is appropriate for the measured continuum flux which is
compact, but the imaging data show the excess line emission to have a shallow
surface brightness profile that will be close to constant within the PSA.  We
apply a 13~\% correction for vignetting of the COS aperture\footnote{COS
Instrument Handbook Section 5.9.}.  

The $\lambda = 1032$~\AA\ line in emission  is unfortunately absorbed in the \hI\
\lya\ transition in the halo of an intervening galaxy at $z=0.05$, but the
blueshifted \oVI\ absorption is still clear and resolved. Further supporting the
detection of \oVI\ emission, we also show the averaged spectrum of 20
star-forming galaxies at similar redshift, which also shows an emission feature
in at 1038~\AA.  More information on these archival spectra is provided in
Section~\ref{sect:archive}. 

The lower redshift absorption system is particularly unfortunate because it
absorbs any emission from the $\lambda 1032$~\AA\ part of the doublet.  The
interloper falls somewhat to the red, but is sufficient to remove the entire
feature.  Fortunately \oVI, as one of the highest ionization potential lines in
the UV spectrum, is significantly blueshifted \citep[see, e.g.][]{Grimes2009}
and the $1032$~\AA\ absorption feature is offset from the interloping absorption
line.  The average blueshift in the local starburst sample observed with FUSE
\citep{Grimes2009} is $-180$~\kms.  In J1156 the SFR is significantly higher,
the galaxy more compact, and the outflow velocity is higher: $-380$~\kms, and
comparable to the maximum in the FUSE sample.  A very similar blueshift of
\oVI~$\lambda 1032$~\AA\ is seen in the stacked spectra of galaxies with similar
SFRs (Section~\ref{sect:archive} and Figure~\ref{fig:archive}).  This blueshift
of the $\lambda 1032$~\AA\ line is particularly fortunate because it enables us
to measure the outflow velocity, despite the lower redshift interloper and the
blending of the $\lambda 1038$~\AA\ line with \cII~$\lambda 1036$~\AA.  

The absorption system may, however, cause us to underestimate the \oVI\ column
density by lowering the apparent continuum near the \oVI\ line.  In this case,
our estimate will be a lower limit, but it is worth noting that J1156 lies well
on the observed relationship between \NoVI\ and velocity width, that unifies all
\oVI\ absorbing systems from the Milky Way disk to IGM systems
\citep{Heckman2002,Grimes2009}. 

\begin{figure*}[t!]
\includegraphics[angle=00,width=18cm]{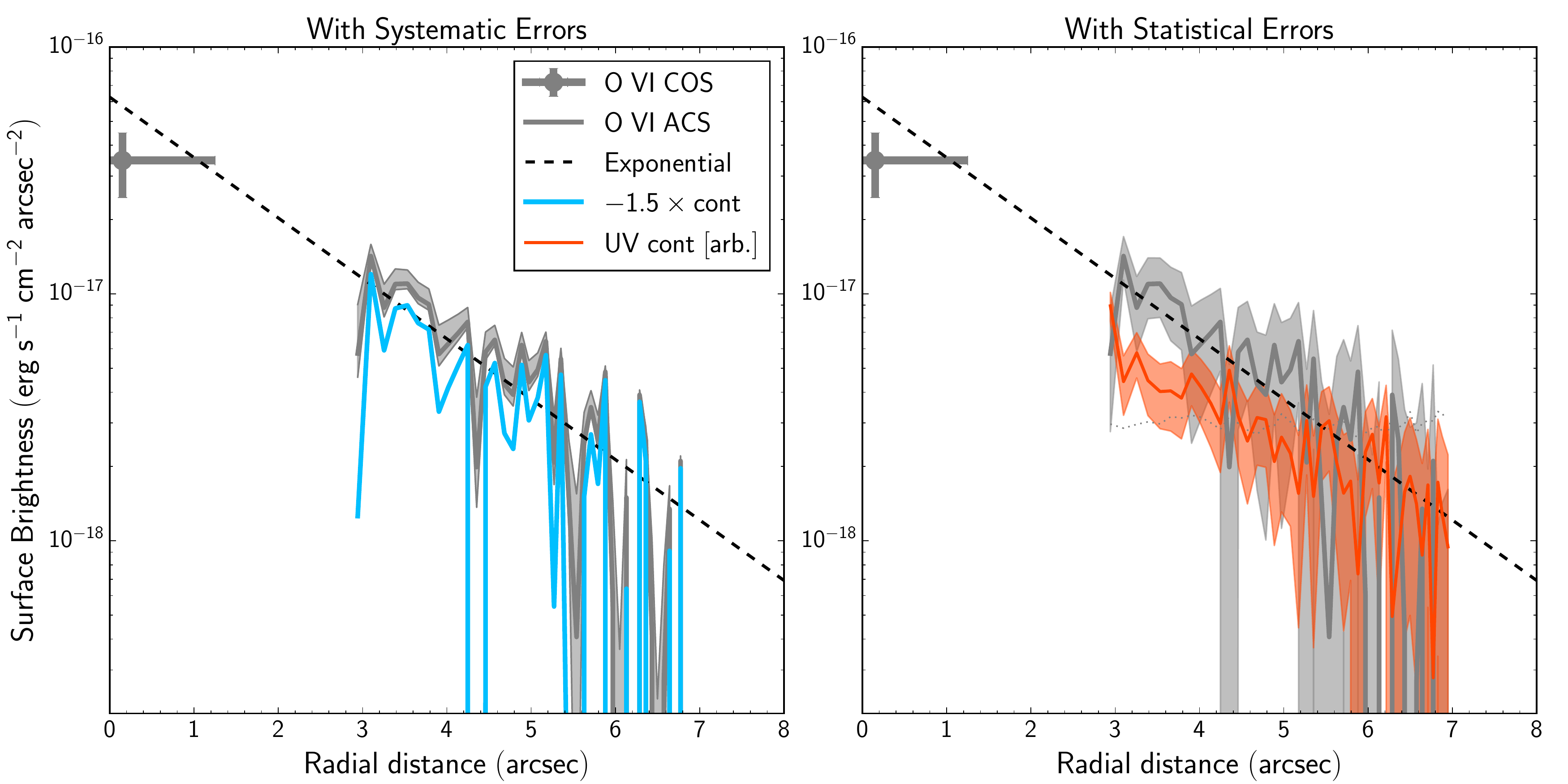}
\caption{The \oVI\ surface brightness profile. The gray point at radius $\approx
0.1$~arcsec is the spectroscopic measurement, slightly offset for visibility, with
x-errorbar showing the radius of the aperture. The line beginning at radius
$\sim 2.5$~arcsec shows the surface brightness measured in the images in isophotal
apertures. Due to the logarithmic scaling, and the fact that more flux is
absorbed in total (mainly in \lyb) only flux beyond $\sim 2$~arcsec can be visualized.
The dashed black line shows the best-fitting exponential profile.  Shaded regions
show the estimated systematic error budget from varying the slope of the UV
continuum (\emph{left}) and upper and lower limits of the 68th percentile statistical
errors from Monte Carlo simulations (\emph{right}).  The \emph{right} panel also
shows the profile of the UV continuum radiation.}\label{fig:radprof}
\end{figure*}

We measure the \oVI\ surface brightness from the continuum-subtracted images,
using isophotally-defined apertures.  This, together with the value measured
centrally by COS spectroscopy, is shown in Figure~\ref{fig:radprof}.  Caution is
needed when comparing this surface brightness profile to the images shown in
Figures~\ref{fig:oviimg} and \ref{fig:methcomp}.  Pixels are averaged over long
annular regions in which the line may be present in either absorption or
emission, and the averaging is performed on pixels that do not have the same
sign.  Thus the annular average may be very much smaller than the peak surface
brightness, while the eye is usually guided by the brightest pixels in an image.
This is especially true for Figure~\ref{fig:oviimg}, which is logarithmically
scaled and the negative pixels are not displayed.  This display issue is common
to all profiles nebular lines where the feature can present in both absorption
and emission.

The central absorption becomes positive at radii beyond $\approx 2.5$~arcsec,
corresponding to 9.4~kpc.  The central surface brightness is higher than at
larger radii, and clearly declines to radii of $\sim 6$~arcsec (23 kpc).  We do
not attempt to constrain the shape of the surface brightness profile but show
that it can be described by an exponential function.  We fit the profile using
the a simple least-squares minimizer, that gives a central surface brightness
$\mu_\mathrm{OVI} = (5.3_{-1.1}^{+1.2}) \times 10^{-17}$~\ergseccmarcsec, with
an exponential scale length of $R_\mathrm{OVI} = 2.0_{-0.2}^{+0.2}$~arcsec.  In
physical distances, this corresponds to $7.5_{-0.7}^{+0.7}$~kpc.  These errors
correspond to the $16^\mathrm{th}$ and $84^\mathrm{th}$ percentiles, determined
by Monte Carlo simulations of the original science frames, which were randomized
using their corresponding weight images; continuum subtractions and profile fits
were computed for each realization.  Integrating this profile to infinity gives
a total \oVI\ flux of $(12.3_{-0.7}^{+1.1}) \times 10^{-16}$~\ergseccm\ and a
luminosity of $(20.5_{-1.2}^{+1.8})\times 10^{40}$~\ergsec.  Because this
measures only the $\lambda 1038$~\AA\ part of the doublet, the total \oVI\
output  may be $\sim 2-3$ times higher \citep{Grimes2007}. These numbers are
summarized in Table~\ref{tab:oviprops}, which we will continue to populate
throughout the following Sections.

\begin{deluxetable*}{rcccl}[t]
\tabletypesize{\scriptsize}
\tablecaption{Derived Properties of \oVI\ and Gas in the Coronal Phase.\label{tab:oviprops}}
\tablewidth{13cm}
\tablehead{
\colhead{Quantity} & \colhead{Symbol used} & \colhead{\oVI} & \colhead{Coronal} & \colhead{Unit}  \\ 
\colhead{(1)} & \colhead{(2)} & \colhead{(3)} & \colhead{(4)} & \colhead{(5)}  }
\startdata
\multicolumn{5}{l}{\textbf{Imaging}}   \\ 
Central Surface Brightness & $\mu_\mathrm{OVI}$ &  $5.3_{-1.1}^{+1.2}\times 10^{-17}$   & \nodata             & \ergseccmarcsec  \\
Angular scale length       & $R_\mathrm{OVI}$   &  $2.0\pm 0.2$                         & \nodata             & arcsec           \\
Physical scale length      & $R_\mathrm{OVI}$   &  $7.5\pm 0.7$                         & \nodata             & kpc              \\
Total flux                 & $f_\mathrm{OVI}$   &  $12.3_{-0.7}^{+1.1} \times 10^{-16}$ & \nodata             & \ergseccm        \\
Total luminosity           & $L_\mathrm{OVI}$   &  $20.5_{-1.2}^{+1.8} \times 10^{40}$  & \nodata             & \ergsec          \\
\hline
\multicolumn{5}{l}{\textbf{Spectroscopy}} \\   
Column density             & \NoVI              & $10^{14.7}$                           & $10^{19.2}$         & \percm           \\
Outflow velocity           & $V_\mathrm{OVI}$   & 380                                   & \nodata             & \kms             \\
\hline
\multicolumn{5}{l}{\textbf{Combination}} \\ 
Density                    & $n$                & $1.7\times10^{-05}$                   & 0.50                & cm$^{-3}$        \\
Column length of clouds    & $D_\mathrm{OVI}$   & 10                                    & \nodata             & pc               \\
Gas mass                   & $M_\mathrm{OVI}$   & $3\times 10^4$                        & $5.5\times 10^7$    & \msun            \\
Cooling time               & $t_\mathrm{cool}$  & 1.3                                   & \nodata             & Myr              \\
Sound crossing time        & $t_\mathrm{sound}$ & $10^4$                                & \nodata             & yr               \\
Total kinetic energy       & $E_K$              & $3.2\times 10^{52}$                   & $5.9\times 10^{55}$ & erg              \\
Momentum                   &                    & $1.7\times 10^{45}$                   & $3.1\times 10^{48}$ & g~cm~s$^{-1}$    \\               
Pressure                   & $p$                & \nodata                               & $1.6\times 10^{5}$  & K~cm$^{-3}$                    
\enddata
\tablecomments{Column (1) lists the quantity, and (2) the symbol used in the
Paper.  Column (3) presents the measured value where appropriate for \oVI. 
Column (4) presents, where appropriate, the same quantity but converted into a 
corresponding value for the mostly hydrogen gas that is assumed to be in CIE with 
the \oVI.  The Table is divided by horizontal lines into quantities that are 
derived from imaging, spectroscopy, and the combination of the two.}
\end{deluxetable*}

\subsection{Recombination Line Emission}

\begin{figure*}[t!]
\includegraphics[angle=00,width=18cm]{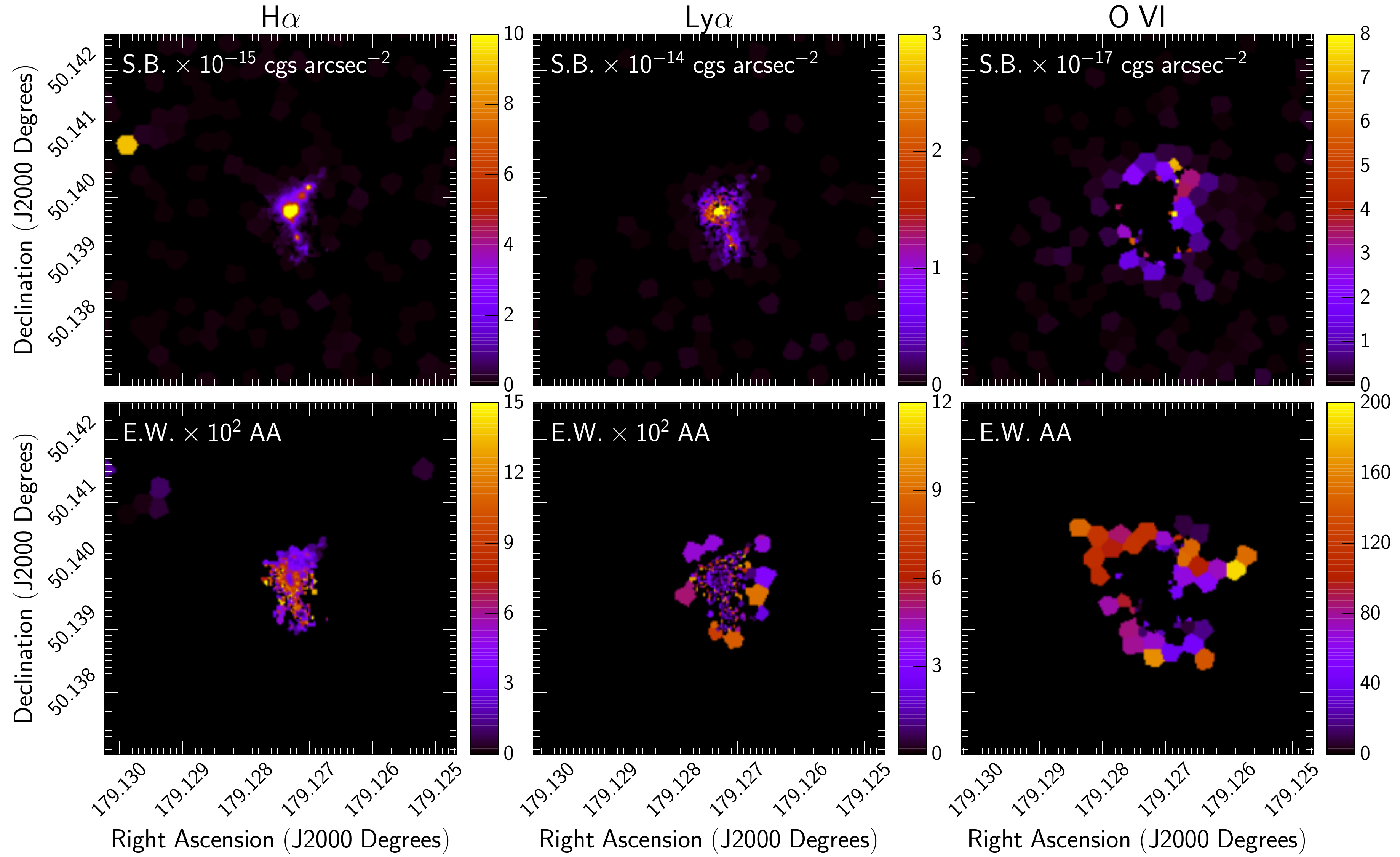}
\caption{J1156 in various continuum-subtracted emission lines.  \emph{Left}
shows \halpha, \emph{centre} shows \lya, and \emph{right} shows \oVI.  Here we
present the Voronoi tessellated data.  Surface brightness levels in
\ergseccmarcsec, and are different in each panel, each of
which has its own color-bar.  }\label{fig:ovilyaha}
\end{figure*}

\begin{figure}[t!]
\includegraphics[angle=00,width=9cm]{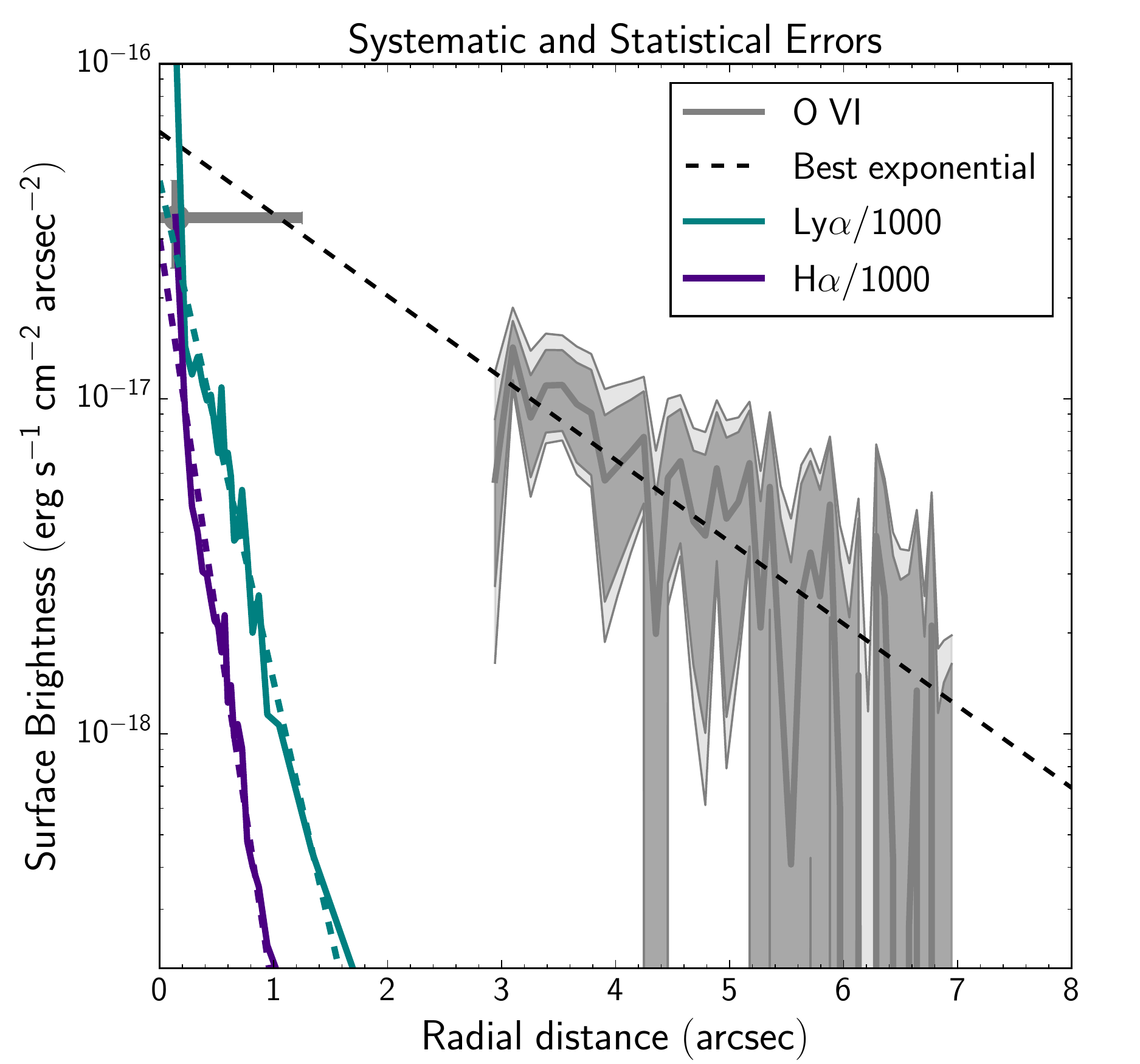}
\caption{The same as Figure~\ref{fig:radprof} except the systematic and
statistical errors have been combined.  Here we also add the radial profiles of
\lya\ (cyan) and \halpha\ (blue) in order to compare with emission that comes
purely from photoionization.  \lya\ and \halpha\ profiles have been divided by
1000 to be included in the same Figure.}\label{fig:radprofcomp}
\end{figure}

In Figure~\ref{fig:ovilyaha} we show the continuum subtracted \lya\ and \halpha\
images, together with \oVI.  Note that different intensity ranges are used in
each panel and, as the Figure shows, the peak \lya\ surface brightness is on the
order of 100 times that of \oVI.  Due to the significantly shorter integration
times in the \lya\ and \halpha\ filters, we are not able to probe the same
regions that we can in \oVI: the F125LP image (\oVI) has over 20 times the
exposure time of F150LP (\lya) and also has around twice the peak throughput
(Figure~\ref{fig:filters}).    The radial profiles of \lya\ and \halpha,
together with that of \oVI, are shown in Figure~\ref{fig:radprofcomp}.  As we
cannot compare line fluxes at large radii, we instead measure the corresponding
exponential scale lengths in \halpha\ and \lya.  In practice it is difficult to
obtain a robust result because of the very compact nature of the starburst.  For
\halpha\ we obtain an exponential scale length of 0.2~arcsec (0.75~kpc) and
$\sim0.3$~arcsec (1.1~kpc) for \lya, where the error is about 0.1~arcsec.
These scale lengths are $\sim 1/10$ and $\sim 1/7$ that of the \oVI\ profile. 

Interestingly the \lya/\halpha\ ratio is high: aperture flux growth curves of
\lya\ and \halpha\ converge at fluxes of $1.1\times 10^{-13}$ and $3.7\times
10^{-14}$~\ergseccm, respectively.  With a \lya\ luminosity of
$1.8\times10^{43}$~\ergsec, J1156 is the most luminous known \lya-emitting
galaxy at $z<0.8$ (\citealt{Cowie2011,Hayes2014,Henry2015,Hayes2015}, c.f.
\citealt{Wold2014}).  A \lya/\halpha\ ratio of $\approx 2.9$ is precisely
what is expected for pure dust attenuation using the measured \halpha/\hbeta\
ratio -- there is little room for significant preferential absorption of \lya\
because of a large number of resonance scatterings.

\subsection{Is the halo \oVI?}

Having isolated a potential flux excess in the on-line filter, we must now
solidly address two questions: \emph{is the excess flux real and significant?}
and, if so, \emph{is it a halo of \oVI\ emission?}

\subsubsection{Instrumental and Methodological Effects}

Turning first to the question of the reality, the most obvious and important
issues are those of instrumental effects such as residual dark current, red
leaks in the filters, and the point spread function (PSF) or scattered light.
ACS/SBC indeed does have a significant dark current that is centrally peaked.
However the dark current is strongly temperature dependent: below $\approx 25$
degrees Celsius the dark current is slightly below $10^{-5}$
counts~sec$^{-1}$~px$^{-1}$, while above this temperature it grows
substantially.  From a cold start the SBC reaches this temperature after
$\approx 3$~hours of operating time, which corresponds to about two orbits.  As
discussed in Section~\ref{sect:obs:imgdata}, we firstly divided our 16 orbits of
F125LP and F140LP observations into eight visits of two orbits, and secondly
arranged with STScI for our visits to be executed only when the SBC has not been
used in the previous 24 hour period.  According to the current status of the
SBC, the dark current should not be detectable in any of our images.  Moreover,
we arranged the exposures within our visits such the F125LP frames are
interleaved between the F140LP in the same orbit, using shadow time for F125LP.
Thus, even if there were some residual dark current in the frames it
should be well balanced between the two exposures. 

The PSF wings of the SBC are particularly broad, and a mis-match between the
F125LP and F140LP filters could mimic such a halo.  Measuring the PSF at large
radii, and at a wavelength where the majority of stars are very faint, is
extremely challenging.  We have made a major effort to characterize the wings of
the SBC PSF in the relevant filters, used here and in the LARS project (Melinder
et al.  in preparation).  The method and main results can be found in
Appendix~\ref{sect:psf} of this article and a much more general study of the HST
PSFs is in preparation.  In short we obtained data of white dwarf stars in the
globular cluster NGC\,6681, drizzled and stacked the images in each filter.  We
then identified all the UV sources in the stacked image, and selected the 20
brightest stars from which to build an empirical PSF.  We extracted stamps of
the individual stars, normalized and stacked them.  To combat contamination from
nearby sources, we then used this 2-dimensional PSF to remove all objects apart
from one from the drizzled image, and thus produced 20 images containing only
one of the 20 reference stars in each.  We then repeated the extraction,
stacking, and removal three times until the empirical PSF converged.  This
procedure enabled us to measure the PSF wings of F125LP and F140LP out to radii
of 3~arcsec, which we show to be indistinguishable at these distances.  We begin
to see halo emission in the continuum-subtracted image at radii somewhat smaller
than this, and even if J1156 were a point source, PSF wings would not be able to
explain the halo.  Since the galaxy is not point-like, PSF issues or scattered
light may not explain the detection, although we cannot rule out minor effects
on the photometry at larger radii still.

\subsubsection{Significance of the Detection}\label{sect:res:sign}

Next arises the question of how significant the detection is.  This depends upon
an accurate understanding of the noise characteristics, the treatment of the
reduced frames, and the continuum subtraction.  

As described in Section~\ref{sect:contsub}, we execute Monte Carlo simulations
to determine the significance of the detection.  Since we drizzle the pixels
from a fine sampling (0.034~\arcsec) to a coarser one (0.1~\arcsec), noise
correlation from individual pixels will be minor, but account for it formally
nonetheless \citep{Casertano2000,Fruchter2002}.  Our Monte Carlo simulations are
conducted end-to-end, and the random numbers that are drawn are used only to
re-realize the science frames.  All subsequent calculations are carried out
using the original method in resampled data, and in the surface brightness
profile (\emph{right} panel of Figure~\ref{fig:radprof}) each pixel error
contributes precisely once per realization.  At radii between 2.5 and $\approx
4.5$~\arcsec\ the annular SNR is around $3$ on average; this radial range
includes 41 spatial elements, and if summed only in this region gives a total
SNR of 21.  Note that this SNR is around a factor of 2 larger than that quoted
for the integrated flux (Table~\ref{tab:oviprops}) when the surface brightness
profile is integrated to infinity.   The reason for this is the large
extrapolation in radius.  The relative error becomes larger beyond $\sim
5$~arcsec, and rapidly the signal becomes indistinguishable from the noise.
Statistical errors, while obviously important, cannot explain our detection.

As suggested in Section~\ref{sect:contsub}, we have performed the continuum
subtraction based upon a single assumption for the UV continuum slope
($\beta=-1.9$), but we also adopt a range of  $-3.4 < \beta < 0$ to explore
systematic errors.  $\beta$ of $-3.4$ is far bluer than stellar models can
reproduce, but is the largest value measured in the $\beta$-map.  Stellar
populations can become arbitrarily red, especially in the UV, but our lower
value corresponds to a slope that is flat in $f_\lambda$.  

These limiting surface brightness profiles are shown by the gray error region in
the left panel of Figure~\ref{fig:radprof}.  It is true that stellar continuum
slopes can become somewhat redder but they cannot become bluer.  Were our
assumed slope too red we would over-predict the line emission, but in adopting
$-3.4$ as the limiting case this is the most pessimistic continuum subtraction
we can envisage, with $\zeta_\mathrm{UV}\approx 1.2$.  The key point in
assessing \oVI\ surface brightness is that redder slopes will only result in
more \oVI.  This point is particularly important: Figure~\ref{fig:beta} shows
that the continuum is rather red in the regions where the \oVI\ emission is
seen: if the \oVI\ were the result of significantly extended PSF wings in F125LP
we would infer a blue slope in these regions, not redder.  This cannot affect
the detection of excess flux. 

While the approach assumes a parametric form of the slope, this encompasses all
feasible variations in stellar metallicity, stellar age, segregation of low and
high mass stars, and reddening by dust.  It is clear from
Figure~\ref{fig:radprof} that while the UV stellar continuum does affect the
\oVI\ surface brightness, the detection itself cannot arise artificially from a
poor estimate of the UV slope.  Finally, we execute tests where we artificially
multiply the offline filter by completely unrealistic $\zeta_\mathrm{UV}$
factors (2, 3, and 5 were tested) and still the result is an excess flux in the
\oVI\ image.  As the right panel of the same Figure shows, the statistical
errors actually far outweigh the systematic ones.

\subsubsection{Sources of the Emission}\label{sect:emsources}

Having argued that the excess flux is real line emission we must establish that
it is indeed \oVI, because other absorption, and potential emission features
fall within the bandpass.  The \lyb\ line will be intrinsically 1/3 the strength
of \lya\ under Case A conditions: \lya\ is shown in Figures~\ref{fig:ovilyaha}
and \ref{fig:radprofcomp} and is hundreds of times brighter than \oVI.  When
\lya\ is emitted by local starbursts, it is almost ubiquitously with the
morphology of large halos \citep[e.g.][]{Hayes2013} due to scattering in atomic
hydrogen.  \lyb\ may also scatter in \hI, but at each event the probability of a
$^3P_{1/2} \rightarrow ^1S_{1/2}$ \lyb\ transition is around $4/5$, while the
remaining $1/5$ is a decay to the $^2S_{1/2}$ level resulting in \halpha, then
two-photon,  emission.  Thus while \lya\ can scatter indefinitely, \lyb\ may do
so only a handful of times.  We may expect \lyb\ in principle to be emitted if
it can escape directly from the central regions, but any halo should not be as
extended as \lya, and in realistic circumstances would exhibit a steeper
profile, if seen at all.  Since we do not see a \lyb\ line in the centre of the
galaxy, and the \lya\ profile declines far more steeply than the suspected \oVI\
halo, we do not believe our halo is one of scattered \lyb.  Moreover the entire
FUSE archive of starbursts \citep{Grimes2009}, and more recent COS observations
of similar galaxies \citep[e.g.][see also
Section~\ref{sect:archive}]{Heckman2011,Henry2015}, do not present convincing
cases for \lyb\ emission, and we anyway do not believe the Case A scenario
needed for \lyb.  \oVI\ on the other hand has, albeit rarely, been seen in
emission.

The \cII~$\lambda1036$~\AA\ feature is also a resonance line and therefore
scatters as well as absorbs, and may also reprocess radiation into a fluorescent
emission line (\cII*) with wavelength 1037.0~\AA.  This fluorescent line, if
present, is ~190~\kms\ bluewards of the \oVI~$\lambda 1038$~\AA\ line, and
indeed a small peak is visible in the spectrum which could possibly be \cII*.
If so, however, this is around $1/5$ the strength of the line we have identified
as \oVI~$\lambda 1038$~\AA, and is therefore unlikely to be the emission we see
at large $R$. 

After eliminating \lyb, scattered \cII, and fluorescent \cII* emission, we
conclude that the excess emission in the F125LP filter arises from the \oVI\
transition at 1038~\AA.  However we do not know whether it is due to gas that is
collisionally ionized (or even photoionized) in situ, or whether it is continuum
radiation that is resonantly scattered by \oVI\ ions.  Indeed if there is gas
there to be emitting because of collisional ionization, it must also be
scattering.  This was discussed in \citet{Grimes2007}, who pointed out that  in
\har\ the emitted \oVI\ was of roughly equal EW to the absorbed line, which
would be expected for scattering.  Examining the COS spectrum of J1156 -- which
samples a similar physical scale to the FUSE in \har\ (see
Section~\ref{sect:lit}) -- the situation appears to be very similar.  However
the emitting region in the images that we interpret as \oVI\ extends over a far
greater area than that covered by the COS aperture, and when the exponential
profile is integrated to infinity the luminosity is about eight times higher
than in the COS aperture alone: $2.5\times 10^{40}$ c.f. $2.0\times
10^{41}$~\ergsec, as further discussed in Section~\ref{sect:lit}.  In contrast,
more than 80~\% of the UV continuum flux, measured in F140LP, comes from within
the COS aperture.  For a spherically symmetric case a component of the emission
must be due to scattered continuum radiation, but we estimate that this
mechanism contributes only $\approx 1/6$ of the total halo emission.  Of course
a departure from spherical symmetry  -- for example if we are looking into a
lower-than-average column of coronal gas -- will increase the scattered
component.  Currently there is no obvious way to estimating this for J1156.

In summary, we believe that while the signal is weak in general, we have
detections of \oVI\ in both imaging and spectroscopic mode.  The signal is far
above the noise based upon both statistical and systematic error budgets, and
the available spectroscopic information suggest that \oVI~$1038$~\AA\ is the
most likely feature.  We argue that only a minority of the measured emission
should be due to resonance scattering.  Furthermore, a very similar feature is
visible in stacked spectra of archival observations, which we review in the next
Section.

\section{Spectra Around \oVI\ for an ensemble of starbursts}\label{sect:archive}

We searched the MAST archive for COS/G130M observations of starburst galaxies
with redshifts that place the \oVI\ doublet in the wavelength region covered by
the grating $(z = 0.12 - 0.4)$.  Two main studies dominate this archive search,
one targeting the nearby analogues of Lyman Break Galaxies -- so-called Lyman
Break Analogs (GO 11727 and GO 13017; PI: T. Heckman) -- and one targeting Green
Pea galaxies (GO 12928; PI: A. Henry).  These galaxies are summarized in
Table~\ref{tab:archive}. Spectra were obtained and processed in an identical
manner as for J1156, as described in Section~\ref{sect:obs:spec}.

\begin{deluxetable*}{lcccccc}[t]
\tabletypesize{\scriptsize}
\tablecaption{Archival galaxies used in the stacking analysis. \label{tab:archive}}
\tablewidth{0pt}
\tablehead{
\colhead{SDSS ID} & \colhead{RA} & \colhead{Dec} & \colhead{Redshift} &
\colhead{Prog. ID} & \colhead{\lha} & \colhead{\ewha} \\ 
\colhead{} & \colhead{J2000} & \colhead{J2000} & \colhead{} &
\colhead{} & \colhead{$10^{42}$\ergsec} & \colhead{\AA} \\ 
\colhead{(1)} & \colhead{(2)} & \colhead{(3)} & \colhead{(4)} & \colhead{(5)} &
\colhead{(6)} & \colhead{(7)} }
\startdata
J005527.46-002148.7 & 00:55:27.5 & $-$00:21:48.7 & 0.16742 & 11727 & 1.76  & 379.2 \\
J015028.4+130858.3  & 01:50:28.4 & $+$13:08:58.3 & 0.14667 & 11727 & 1.03  & 200.6 \\
J030321.41-075923.2 & 03:03:21.4 & $-$07:59:23.2 & 0.16481 & 12928 & 1.18  & 589.8 \\
J091113.34+183108.1 & 09:11:13.3 & $+$18:31:08.1 & 0.26217 & 12928 & 1.90  & 390.3 \\
J092159.38+450912.3 & 09:21:59.4 & $+$45:09:12.3 & 0.23497 & 11727 & 1.08  & 74.24 \\
J092600.4+442736.1  & 09:26:00.4 & $+$44:27:36.1 & 0.18067 & 11727 & 1.41  & 578.0 \\
J105330.82+523752.8 & 10:53:30.8 & $+$52:37:52.8 & 0.25260 & 12928 & 2.50  & 375.6 \\
J111244.05+550347.1 & 11:12:44.0 & $+$55:03:47.1 & 0.13163 & 13017 & 1.43  & 205.8 \\
J111323.99+293039.2 & 11:13:23.9 & $+$29:30:39.2 & 0.17514 & 13017 & 0.104 & 26.30 \\
J113303.78+651341.3 & 11:33:03.8 & $+$65:13:41.3 & 0.24141 & 12928 & 0.649 & 277.1 \\
J113722.13+352426.6 & 11:37:22.1 & $+$35:24:26.6 & 0.19431 & 12928 & 2.03  & 562.8 \\
J114422.31+401221.2 & 11:44:22.3 & $+$40:12:21.2 & 0.12695 & 13017 & 0.443 & 87.38 \\
J121903.98+152608.5 & 12:19:04.0 & $+$15:26:08.5 & 0.19558 & 12928 & 1.63  & 1094 \\
J124423.37+021540.4 & 12:44:23.4 & $+$02:15:40.4 & 0.23938 & 12928 & 3.15  & 777.5 \\
J124834.63+123402.9 & 12:48:34.6 & $+$12:34:02.9 & 0.26341 & 12928 & 1.64  & 669.5 \\
J141612.96+122340.5 & 14:16:12.9 & $+$12:23:40.5 & 0.12316 & 13017 & 1.11  & 184.5 \\
J142405.72+421646.2 & 14:24:05.7 & $+$42:16:46.2 & 0.18482 & 12928 & 2.14  & 1083 \\
J142856.4+165339.4  & 14:28:56.4 & $+$16:53:39.4 & 0.18167 & 13017 & 1.39  & 251.2 \\
J161245.59+081701   & 16:12:45.5 & $+$08:17:01.0 & 0.14914 & 13017 & 1.62  & 175.4 \\
J210358.74-072802.3 & 21:03:58.8 & $-$07:28:02.3 & 0.13677 & 11727 & 1.95  & 107.3 \\
\hline
J115630.63+500822.1 & 11:56:30.4 & $+$50:08:28.0 & 0.23599 & 13656 & 3.32  & 369.4
\enddata
\tablecomments{Galaxies are ordered by right ascension.  J1156 is not included
in the stacking analysis but is included for comparative purposes in the last
row. }
\end{deluxetable*}

To examine the \oVI\ feature in an ensemble of galaxies, we averaged the spectra
in a number of sub-samples.  We first measured the systemic redshifts by
obtaining the SDSS spectrum, and measuring the wavelength centroid of the
\halpha, \hbeta, and [\oIII]~$\lambda \lambda 4959,5007$~\AA\ lines.  We
converted each to a redshift and took the unweighted average for each galaxy.
We then produced a mask for each object to exclude wavelength regions where the
continuum is affected by interstellar absorption lines (in both the Milky Way
and the target galaxy) and stellar features (in the target galaxy only).  We
fitted a low order polynomial function to each spectrum and normalized each one.
We then blueshifted each spectrum into the restframe using the redshift obtained
from nebular lines.  With the spectra normalized and shifted into the restframe,
we finally masked the Milky Way absorption features in each spectrum and
computed a simple average.  We produced stacks for the full sample of 20
galaxies, and also for subsamples by dividing the sample in two halves when
sorted by \halpha\ both luminosity and equivalent width.

\begin{figure*}[t!]
\includegraphics[angle=00,width=18cm]{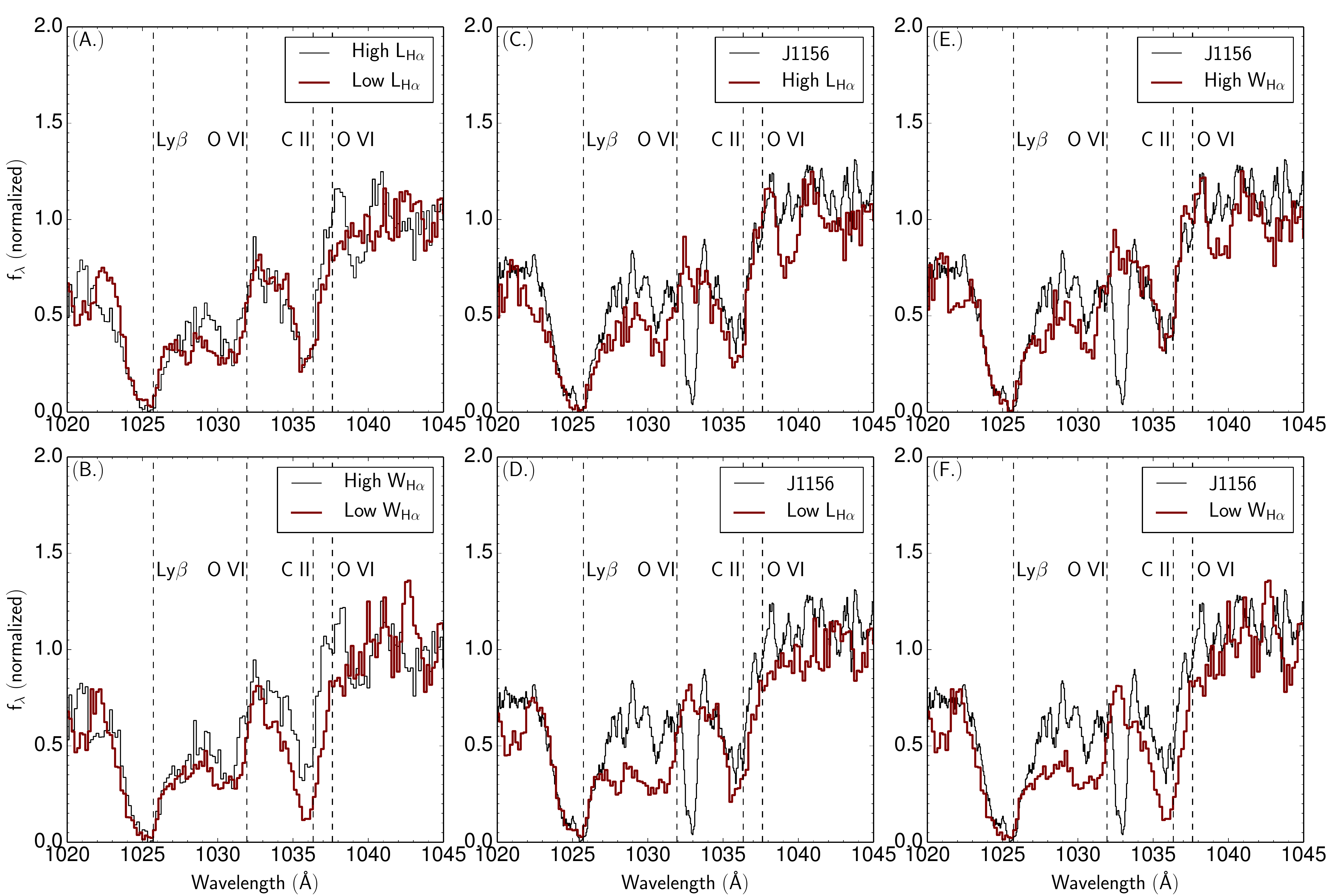}
\caption{The Differential Stacking Analysis.  The \emph{left} panels show the
comparison of the higher and lower \lha\ subsamples (\emph{upper}), and the
higher and lower \ewha\ subsamples (\emph{lower}).  The \emph{central} and
\emph{right} panels compare the high/low \lha/\ewha\ subsamples with J1156.  The
wavelengths of \lyb, \cII~$\lambda1036$~\AA, and the \oVI\ doublet are
marked.}\label{fig:archive}
\end{figure*}

The average properties of the galaxies in the stacks are presented in
Table~\ref{tab:archstat} and the resulting spectra are shown in
Figure~\ref{fig:archive}.  We briefly compare the observables of J1156 with
those of the stacked samples.  J1156 has a \halpha\ luminosity of $3.3\times
10^{42}$~\ergsec\ (SFR = 26~\msunyr; \citealt{Kennicutt1998}).  This is
marginally higher than the most strongly star-forming galaxy in the archival
sample (25~\msunyr).  The high \ewha\ stack has \ewha\ in the range
375--1094~\AA, while J1156 shows \ewha\ of 370~\AA, and would thus be the median
galaxy if included in the stack. 

We first compare the intensely star-forming subsamples with their more weakly
star-forming counterparts in the \emph{left} panel of Figure~\ref{fig:archive}.
The absorption regions of the spectrum appear broadly similar.  This is true
particularly for \lyb, is true also for the blended \cII~$\lambda 1036$~\AA\ and
\oVI~$\lambda 1038$~\AA\ lines, and to a lesser extent holds for the
\oVI~$\lambda 1032$~\AA\ feature.  The most striking difference that emerges
when intensely star-forming galaxies are isolated in the stack is the presence
of an emission feature at $\lambda\approx 1038$~\AA\ that does not appear in the
stacks of less star-forming galaxies.  In the less star-forming galaxies the
continuum rolls more smoothly into the absorption feature without discrete
feature.  

We have investigated the possibility that the apparent increased strength of the
1038~\AA\ emission may instead result from stronger \oI~1039~\AA\ absorption,
that produces a spurious emission feature.  The archival spectra also extend to
redder wavelengths and include the \oI~1302~\AA\ electric dipole transition.
These two \oI\ features arise from the same lower energy level and the two lines
are expected to covary.   In fact, the transition probability for 1302~\AA\ is
$\approx 7$ times higher and we can therefore use it as a test of what to expect
at 1039~\AA.  Unlike at 1039~\AA, the 1302~\AA\ line is clearly visible in all
sub-stacks, and we measure EWs in the range 0.86--1.12~\AA.  Indeed the
absorption is slightly stronger in the higher SFR sub-stacks, but the relative
difference is only $\approx 25$~\%.  Typical errors in EW are around 0.2~\AA,
and this difference is barely significant.  Moreover even if the trend were to
hold the magnitude is too small to explain the difference at 1039~\AA.  Given
that the 1302~\AA\ lines are present in all stacked spectra, we do not believe
the 1038~\AA\ emission feature arises purely from increased \oI\ absorption at
1039~\AA, and the most straightforward interpretation is that the peak is \oVI\
emission in the higher SFR and higher \ewha\ galaxies.

The \emph{centre} and \emph{right} panels of Figure~\ref{fig:archive} show the
comparison of the J1156 spectrum with stacks of the low and high \lha\ and
\ewha\ subsamples.  Both of the intensely star-forming stacks show remarkable
similarity to J1156.  For example, the shape and offset of the \lyb\ are almost
indistinguishable between J1156 and the average of 10 starbursts.  Similarly the
\cII~$\lambda 1036$~\AA\ cannot easily be separated by eye.  Most importantly,
and as in the previous paragraph, the \oVI~$\lambda 1038$~\AA\ feature appears
similar to the stacked spectra in the two strongly star-forming cases: it forms
at the same rest wavelength with a similar equivalent width (as well as EW can
be measured).  Note that it is possible that both emission and absorption lines
are likely to appear artificially broadened in a stacking analysis, as they may
not be formed at precisely the same velocity, compared to the optical line
emission.  The comparison supports a scenario in which strongly star-forming
galaxies produce a weak emission feature in their spectrum at the wavelength
where \oVI\ would be expected, and where an excess is seen in J1156.

\begin{deluxetable*}{llcccc}[t]
\tabletypesize{\scriptsize}
\tablecaption{Statistics for the Subsamples.\label{tab:archstat}}
\tablewidth{0pt}
\tablehead{
\colhead{Subsample} & \colhead{Statistic} & \colhead{Redshift} & \colhead{\lha} &
\colhead{\ewha} & \colhead{SFR(\halpha)}  \\ 
\colhead{} & \colhead{} & \colhead{} & \colhead{$10^{42}$\ergsec} &
\colhead{\AA} & \colhead{\msunyr}  \\ 
\colhead{(1)} & \colhead{(2)} & \colhead{(3)} & \colhead{(4)} & \colhead{(5)} &
\colhead{(6)}  }
\startdata
Full        & Mean    & 0.18763 & 1.51  & 404.5 & 11.9 \\
            & Median  & 0.18117 & 1.53  & 326.4 & 12.1 \\
            & Minimum & 0.12316 & 0.10  & 26.30 & 0.82 \\
            & Maximum & 0.26341 & 3.15  & 1094  & 24.9 \\
\hline
High \lha\  & Mean    & 0.20456 & 2.03  & 561.5 & 16.1 \\
            & Median  & 0.19495 & 1.93  & 476.6 & 15.2 \\
            & Minimum & 0.13677 & 1.62  & 107.3 & 12.8 \\
            & Maximum & 0.26341 & 3.15  & 1094  & 24.9 \\
\hline
Low \lha\   & Mean    & 0.17071 & 0.983 & 247.5 & 7.77 \\
            & Median  & 0.16998 & 1.09  & 203.2 & 8.66 \\
            & Minimum & 0.12316 & 0.104 & 26.30 & 0.82 \\
            & Maximum & 0.24141 & 1.43  & 589.8 & 11.3 \\
\hline
High \ewha\ & Mean    & 0.21052 & 1.93  & 650.0 & 15.3 \\
            & Median  & 0.19495 & 1.83  & 583.9 & 14.5 \\
            & Minimum & 0.16481 & 1.18  & 375.6 & 9.33 \\
            & Maximum & 0.26341 & 3.15  & 1094  & 24.9 \\
\hline
Low \ewha\  & Mean    & 0.16475 & 1.08  & 159.0 & 8.54 \\
            & Median  & 0.14791 & 1.10  & 180.0 & 8.66 \\
            & Minimum & 0.12316 & 0.104 & 26.30 & 0.82 \\
            & Maximum & 0.24141 & 1.95  & 277.1 & 15.4
\enddata
\tablecomments{SFRs are lower limits, as they assume no obscuration of \halpha.}
\end{deluxetable*}

\section{Interpreting the \oVI\ Features}\label{sect:interp}

Absorption line studies can target \oVI\ at much larger radial distances than
imaging can, and have been performed out to distances of 300~kpc
\citep{Prochaska2011ovi}.  The \emph{COS-Halos} study has revealed two relevant
correlations: (a.) the decrease of \oVI\ column density (\NoVI) with increasing
impact factor; (b.) the increase in \NoVI\ with star formation intensity.
However these samples contain only two star-forming galaxies with quasars
sightlines inside the 23~kpc over which we see emission.  Our study probes much
smaller distances, in a galaxy that is forming stars 20 times more intensely
than the strongest in \emph{COS-Halos}.  We regard the imaging data to be highly
complementary to halo spectroscopy in that it provides two-dimensional mapping
and probes gas nearer the star-forming region.  Indeed relying upon high volume
density for emission, this is likely to be the only gas that \oVI\ imaging will
be sensitive to.  In the following Sections we derive some physical properties
of the halo gas that we observe, including its density, mass, how it was
excited, and what its fate will be. 

At the end of Section~\ref{sect:emsources} we discussed the possibility that
some of the halo emission could be scattered in the \oVI\ transition and not be
due to collisional processes.  We estimated that in a spherically symmetric
situation only $\sim 1/6$ of the observed \oVI\ would be due to scattering, but
acknowledge that departure from spherical symmetry could lead to a larger
scattered contribution.  We assume that collisional processes dominate in the
calculations that follow, but note that the derived numbers are accurate only in
the case that the scattered contribution is relatively minor.

\subsection{Physical conditions in the \oVI-bearing gas} 

We calculate (or place constraints upon) some of the main physical conditions in
the \oVI\ gas.  Using the apparent optical depth method
\citep{Savage1991,Sembach2003} and the \oVI~$\lambda 1032$~\AA\ absorption line,
we determine the column density of \oVI\ ions to be
$\log(N_\mathrm{OVI}/\mathrm{cm}^{-2}) \approx 14.7$ along the line of sight to
the central star-forming cluster.  The error on this value is about 40~\%, which
dominates the error on all subsequent calculations.  In the same measurement we
obtain a FWHM of 220~\kms; this corresponds to a Doppler parameter $b
(=\sigma\sqrt{2})$ of 132~\kms, and the galaxy lies perfectly among the
starbursting population in the \NoVI--$b$ diagram in \citet{Grimes2009}.  This
diagram, when populated with \oVI\ absorbing systems from the plane of the Milky
Way, galaxy disks, and IGM clouds has been used by \citet{Heckman2002} to
conclude that all \oVI\ absorbing systems have a common physical origin: that
they are cooling flows as they pass through the $T\sim 10^{5.5}$~K phase. 

Assuming the \oVI\ gas we see in absorption is the same as in emission, we may
use the measured column density and a few motivated assumptions to predict the
surface brightness in emission.  To compute the emission we need the normalized
cooling rate, $\Lambda_\mathrm{n}n_\mathrm{e}^2$, and the electron density
$n_\mathrm{e}$, where the cooling coefficient is normally calculated at the
solar oxygen abundance \citep[e.g.][]{Sutherland1993}.  When scaled by the gas
metallicity, these give us the volumetric cooling rate $\Lambda$, in units of
\ergsec~cm$^{-3}$.  With a dimension of luminosities per volume, we may compute
the expected surface brightness if we know the optical length, $D_\mathrm{OVI}$
-- the physical ``thickness" of the emitting and absorbing gas along the
line-of-sight.  That is: 
\begin{equation}
\mu_\mathrm{OVI} = \Lambda_\mathrm{n} n_\mathrm{e}^2
\bigg(\frac{Z}{Z_\odot}\bigg) D_\mathrm{OVI} \label{eq:ovisb}
\end{equation}
where $\mu_\mathrm{OVI}$ is the \oVI\ surface brightness and we have introduced
$Z/Z_\odot$ to scale down the emissivity for less metallic plasma.  While we do
not know $n_\mathrm{e}$ or $D_\mathrm{OVI}$, they are not independent and are
related through the column density by $n_\mathrm{e} =
N_\mathrm{OVI}/[D_\mathrm{OVI}(\mathrm{O/H}) ]$, where (O/H) is the number of
oxygen particle per proton that is in equilibrium with the \oVI.  If we insert
this into Equation~\ref{eq:ovisb} and equate $\mu_\mathrm{OVI}$ to the observed
central surface brightness $\mu_\mathrm{OVI,0}^\mathrm{obs}$ measured in
Section~\ref{sect:res:obs}, we can solve for $D_\mathrm{OVI}$ and obtain: 
\begin{equation} 
D_\mathrm{OVI} = 2\Lambda_\mathrm{n}
\bigg(\frac{N_\mathrm{OVI}}{\mathrm{O/H}}\bigg)^2 \bigg(\frac{Z}{Z_\odot}\bigg)
\frac{1}{\mu_\mathrm{OVI,0}}
\end{equation} 
The factor 2 arises because of the assumption of spherical symmetry where we
integrate over the full column and not the half column from which the column
density is measured (there is as much gas behind the cluster as in front).  Note
also that two metallicity-like quantities enter here: the (O/H) scales the
measured \oVI\ column density to $N_\mathrm{e}$, and the $(Z/Z_\odot)$ scales
the cooling function to the metallicity of the galaxy.  Cancelling one (O/H)
would introduce additional factors. 

We assume the gas is in collisional ionization equilibrium (CIE).  We appreciate
that the gas could in principle be photoionized \citep[e.g.][]{Tripp2008} we
motivate the CIE assumption by (a.) the relative values of \NoVI\ and the
absorption line width $b$, following \citet{Heckman2002}, and (b.) the
fact that the \oVI\ surface brightness profile is ten times more extended than
that of \halpha, which is almost certainly photoionized.  While the propagation
of ionizing photons to these distances cannot be ruled out, photoionization to
\oVII\ would require 138~eV photons and the radiation field would also be rich
with \hI-ionizing radiation as well.  No strong \heII\ emission line is seen in
the optical at 4686~\AA.  For the normalized cooling coefficient in CIE we adopt
that of \citet{Bertone2010}, which is computed for solar metallicity: $\log
(\Lambda/n_\mathrm{e}^2 \mathrm{[erg~s^{-1}~cm^{-3}}]) = -22.5$.  For the
metallicity we use a value of $12+\log(\mathrm{O/H}) = 7.87$, which we derived
from the temperature-sensitive method in Section~\ref{sect:target}.  

The resulting \oVI\ column length is $D_\mathrm{OVI} =10$~pc.   We compare
$D_\mathrm{OVI}$ to the exponential scale length of the \oVI\ emitting surface
(7.5~kpc; Figure~\ref{fig:radprof}): assuming spherical symmetry and that the
absorbing and emitting gas are cospatial and uniformly fill the spectroscopic
aperture, the gas must be strongly clumped, occupying $\approx 1/750$ of the
volume along the line-of-sight ($D_\mathrm{OVI}$ divided by the exponential
scale length).  

A narrow \oVI\ emitting region -- thin with respect to the galaxy halo -- is to
be expected if the $T\approx 10^{5.5}$~K regions represent shocked gas.
\citet{Heckman2001ovi} present hydrodynamical simulations that aim to interpret
the \oVI\ absorption in NGC\,1705, showing \oVI\ to arise only in the
hydrodynamical interaction between hot, out-rushing gas: their Figure~4 is
particularly illustrative in this context, as it shows only the \oVI\ phase, and
would represent a tiny fraction of the size of the object over most sightlines. 

The unique combination of emission and absorption observations outlined above
enable us also to derive the electron density in the coronal phase, giving
$n_\mathrm{e} = 0.50$~cm$^{-3}$.  From this we may also derive the thermal
pressure, for which we obtain $P/k_\mathrm{B} \approx 1.6\times
10^5$~K~cm$^{-3}$, where $k_\mathrm{B}$ is Boltzmann's constant.  At this point
we note the similarity between this number and the value derived in the nebular
phase from the ratio of components in the [\sII] doublet: the ratio of the
6717/6731~\AA\ [\sII] lines actually falls slightly above the low-density
asymptotic value in the electron density the curve \citep{Osterbrock1989}, where
the minimum density is $\approx 4.65$~cm$^{-3}$.  However at $1~\sigma$ our
upper limit on the density is 44~cm$^{-3}$.  From these measurements we obtain
best value for the pressure in the nebular regions of $P/k_\mathrm{B} \approx
6.8\times 10^4$~K~cm$^{-3}$, with $65.4\times 10^4$~K~cm$^{-3}$ as the upper
limit.  This range of pressures closely brackets that measured in the coronal
phase, and raises the question of whether the coronal gas and \hII\ regions
could be in pressure equilibrium. 

If the clouds are not confined by external pressure then they will disrupt on a
few sound crossing times.  From these characteristics of the coronal phase and
the physical thickness of the clouds reported above, we can calculate the
minimum sound crossing time, which is $\approx 10^4$~years.  The computed
crossing time is two orders of magnitude smaller than the cooling time (derived
in the following subsection), and if they were not confined by external pressure
the clouds would likely be dispersed before they are visible through their own
emission.  Thus it seems plausible that the two gas phases are indeed in
pressure equilibrium, although we must admit that the similarity in
$P/k_\mathrm{B}$ could be coincidental.  If a causal connection is correct, we
could be seeing the effects of gas clouds being pressurized from the outside by
the volume-filling hot gas that would be seen at X-ray energies. 

In NGC\,1705, \citet{Heckman2001ovi} compute temperatures and pressures in the
\oVI\ phase, under the assumption that the [\sII]-derived electron density can
be applied to the coronal phase.  In J1156 we find tentative support for this
correspondence, using independent estimates in the warm gas.  The calculations
above are quite approximate, but one should note that the number of assumptions
is actually rather small.  The metallicity cannot be modified by large factors,
and no ionization correction has (yet) entered the calculation because we have
so far only only concerned ourselves with the electrons in collisional
equilibrium with the \oVI.  Indeed the largest assumption is that the gas is in
CIE \citep{Oppenheimer2016}.

\subsection{Mass, Energy, and Momentum in the Coronal Phase}

If we assume that the radial decline in \oVI\ surface brightness is purely a
consequence of the geometrically decreasing column density, then we may compute
the total number of \oVI\ ions in the halo.  This follows from taking the
central \oVI\ column density and the scale length, and integrating the
exponential profile to infinity.  This calculation places just $3\times
10^{4}$~\msun\ in oxygen at $T\approx 10^{5.5}$~K, and $5.5\times 10^{7}$~\msun\
in the coronal phase in CIE with the \oVI.  

We may also compute the nebular gas mass as 
\begin{equation}
M_\mathrm{ion} = \frac{\mu ~ m_\mathrm{p} ~ n_\mathrm{p} ~
L_{\mathrm{H}\beta,0}}{ h\nu_\mathrm{H\beta} ~ n_\mathrm{e} ~
\alpha^\mathrm{eff}_{\mathrm{H}\beta}} 
\end{equation}
where $\mu$ is the mean molecular weight, $m_\mathrm{p}$ and $n_\mathrm{p}$ are
the mass and number density of protons, $L_{\mathrm{H}\beta,0}$ is the intrinsic
\hbeta\ luminosity corrected for dust reddening with \halpha/\hbeta, $h$ is
Planck's constant, $\nu_\mathrm{H\beta}$ is the frequency of \hbeta, and
$\alpha^\mathrm{eff}_{\mathrm{H}\beta}$ is the Case B recombination coefficient
for \hbeta.   Adopting $\mu= 0.5$ for a pure electron-proton gas,
$n_\mathrm{e}=4.65$~cm$^{-3}$ from the [\sII] doublet, and
$\alpha^\mathrm{eff}_{\mathrm{H}\beta}=2\times10^{-14}$~cm$^3$~s$^{-1}$
\citep[][for $T\approx 15,000$K]{Osterbrock1989}, we compute $M_\mathrm{ion}
\approx 2.5\times10^{9}$~\msun.  The corresponding mass in oxygen is $1.3\times
10^{6}$~\msun.   For the \oVI\ mass quoted above, there is 45 times as much mass
in the nebular phase as there is in the coronal. 

If we assume that all of the coronal gas is moving at the characteristic outflow
velocity measured from the \oVI~$\lambda 1032$~\AA\ line, we can compute the
kinetic energy and momentum in the coronal gas.  The kinetic energy in \oVI\
ions alone is $3.2\times 10^{52}$~erg, and $5.9\times 10^{55}$~erg in the whole
coronal phase of the wind.  The momentum in the \oVI\ ions is $1.7\times
10^{45}$~g~cm~s$^{-1}$, and $3.1\times 10^{48}$~g~cm~s$^{-1}$ in the whole
coronal phase of the wind.

We may also estimate the total mechanical energy that has been returned to the
ISM by feedback from star formation. In Section~\ref{sect:target} we performed
SED fitting to constrain the age of the stellar population, arriving at a
best-fitting age of 11.5~Myr, with a $1\sigma$ range of 10--39~Myr.  In the same
calculation, the most frequently recovered star-formation history was an
exponentially declining SFR with $e-$folding time of just 5~Myr.  With the age
so close to the characteristic SFR decay time, constraining the true SFH between
limits of instantaneous and constant is close to impossible.  The \halpha\ EW
also places some constraints: a value of 370~\AA\ is compatible with a lower
limit on the age of just 4.5~Myr \citep{Leitherer1999} in the case of an
instantaneous burst.  Taking the best-fitting values of age=11.5~Myr, stellar
mass of $1.5\times 10^9$~\msun, the total mechanical energy budget, summing over
O star winds and supernova ejecta, is $7.8\times 10^{57}$~erg.  While these
numbers carry a large number of uncertainties and model-dependencies, we
calculate that $\sim 1$~\% of the total available mechanical energy is to be
found in the kinetic energy of the current coronal outflow.  How much
\emph{should} be in this phase at any given time depends upon how long-lived the
\oVI\ phase is, which we address in the next Section.

\subsection{The Fate of the Coronal Gas}

In the previous subsection we derived the cooling rate of the gas, $\Lambda$.
Having also obtained $n_\mathrm{e}$ we may also compute the internal energy $U$,
and the ratio $U/\Lambda$ therefore gives the cooling time, $t_\mathrm{cool} =
1.3$~Myr.  This corresponds to a mass cooling rate of 42~\msunyr (almost
identical to the SFR) and a cooling distance of 0.5~kpc at the outflow velocity
of 380~\kms.  Thus it is clear that the $10^{5.5}$~K gas cannot have traveled
far at this temperature and certainly cannot have been lifted to the scale
height from the nuclear star-forming regions.  Instead the coronal phase must be
constantly replenished. 

The stellar mass of J1156 is $1.5\times 10^{9}$~\msun; at the scale height of
7.5~kpc the escape velocity $v_\mathrm{esc}$, assuming only the mass in stars,
is $\approx 38.6$~\kms.  Thus if stars were the only mass available to bind the
wind, the average velocity would be ten times the escape velocity.  Given that
$v_\mathrm{esc} \propto \sqrt{M}$, the dark matter mass inside of 7.5~kpc must
exceed 100 times the stellar mass in order to prevent the wind from escaping.
The halo abundance-matching analysis of \citet{Moster2013}  suggests that a
galaxy of $M_ \mathrm{stell} = ~1.5\times 10^9$~\msun\ will have a dark matter
halo mass of $\approx~10^{11}$~\msun.  Assuming the dark matter density profile
to be isothermal, $v_\mathrm{esc}$ is approximately 3 times the maximum halo
rotation velocity, $v_\mathrm{circ}$ \citep[e.g.][]{Weiner2009}.
\citet{Maller2004} have shown that for a halo mass of $10^{11}$~\msun,
$v_\mathrm{circ} \approx 65$~\kms.  The corresponding $v_\mathrm{esc}$ is
200~\kms, and the conclusion that the coronal gas is becoming unbound appears
robust assuming that no additional energy is drained from the wind.  

Even if the current outflow velocity exceeds the escape velocity, it does not
necessarily imply that the gas will escape as energy losses from radiation could
play a significant role in the future of the outflow.  Indeed the cooling
distance is just half a kpc, so the gas cools almost in situ.  We calculate that
$\sim1$~\% of the available mechanical energy is accounted for in the coronal
outflow, and the remainder is either still available in the piston gas, will
have cooled already, or was radiated immediately without ever contributing to
the wind fluid. Assuming a constant \oVI\ luminosity with time,
$7.4\times10^{55}$~erg have been radiated in \oVI\ over the 11.5~Myr duration of
the star formation episode.  This is remarkably close to the current kinetic
energy in the wind ($5.9\times10^{55}$~erg, above) but compared to the
$7.8\times10^{57}$~erg returned by supernovae and O stars, this is a minor loss
of energy compared to the total amount that has been available. 

The amount of energy that is truly available in the wind is that supplied
directly to the piston gas, which is not simply the sum of the mechanical energy
returned from stellar winds and SNe.  The mechanical energy in this $T\sim
10^{7.8}$~K wind fluid is dependent upon the thermalization efficiency of the SN
ejecta, commonly denoted $\epsilon$, which is not known in this case (or many
cases).  \citet{Strickland2009} derive $0.3<\epsilon<1$ for nearby starburst
M82, but some simulations suggest that early radiative losses in the densest
cores of strong nuclear starbursts may reduce $\epsilon$ to below 0.1
\citep{Thornton1998,Stevens2003}.  Even if we were to adopt $\epsilon=0.1$, then
only around one tenth of the energy budget would have been radiated through the
\oVI\ emission line.  X-ray observations would be particularly useful in
determining the total energy budget.  

If the gas is ejected from the galaxy, the metals will no longer contribute to
the metal abundance of J1156.  However for the calculated cooling time the gas
will have cooled and recombined within $\approx 1$~Myr and within just 0.5~kpc.
Hence even if it does escape, this gas will be easily detectable through \hI\
\lya\ absorption and transitions of \oI\ in the ultraviolet.

\section{Comparison with Other Detections of \oVI\ Emission}\label{sect:lit}

Finally we compare our J1156 observations with the other two known \oVI-emitting
galaxies that were observed with the FUSE: NGC\,4631 \citep{Otte2003} and \har\
\citep{Grimes2007}.  First, however, a brief description of the different nature
of these observations is warranted.  COS and FUSE are both aperture
spectrographs, but have very different entrance windows: the PSA of COS is a
circular aperture with a diameter of 2.5~arcsec, while FUSE sampled a square of
30~arcsec on the side.  However the fact that the targeted galaxies lie at very
different distances almost entirely compensates for this.  J1156 and the stacked
sample discussed in this paper reside at $z \approx 0.24$ and $\langle z\rangle =
0.19$, respectively ($\approx 1$~Gpc).  At these distances, the COS aperture
samples 9.4~kpc in J1156 and 7.9~kpc in the archival sample.

\har\ lies at 84~Mpc, which is just 1/12 the distance of our COS samples, and
the larger FUSE aperture still subtends 12~kpc (full diameter).  This is just
25~\% larger than the physical size sampled by COS in J1156.  For \har\ the FUSE
aperture captures the entirety of the UV continuum.  The \oVI~$\lambda
1038$~\AA\ flux in the FUSE aperture is $(1.4\pm 0.5) \times
10^{-14}$~\ergseccm, which corresponds to a luminosity of $1.26\times
10^{40}$~\ergsec.  The $\lambda 1038$~\AA\ flux of J1156, measured in the COS
aperture alone, gives a luminosity of $2.5 \times 10^{40}$~\ergsec\ -- note that
this refers to the COS aperture only, and is not the luminosity integrated to
infinity quoted in Table~\ref{tab:oviprops}.  Thus the \oVI\ luminosity of J1156
is within a factor of two that of \har.  Given the very similar natures of the
galaxies, and the $\sim 3\sigma$ detections obtained in both cases, these are
very similar.  In fact, simply scaling the luminosities by the SFRs of the
galaxies would completely reconcile the luminosities.  When we integrate the
exponential surface brightness profile to infinity we obtain a luminosity of $2
\times 10^{41}$\ergsec, which is eight times higher than measured in the COS
aperture.  \har\ is also a particularly compact galaxy, and we may speculate
that the aperture losses may be similar. Note again the FUSE aperture did
not allow \citet{Grimes2007} to distinguish between collisional ionization and
scattering, and it is unfortunate that \har\ is too low redshift to be observed
with our imaging method.

NGC\,4631 is a much more regular edge-on disk galaxy, and is also significantly
nearer, at a distance of just 7.5~Mpc (J1156 is 160 times more distant).  At
this distance FUSE samples just 1.1~kpc.  NGC\,4631 was observed on two
pointings, that targeted an extra-planar X-ray bubble.  Pointing A was obtained
at a disk scale height of 4.8~kpc, while B was positioned 2.5~kpc above the
disk. This type of observation is advantageous because observing away from the
disk reduces contamination from a noisy UV continuum where absorption may blend
with emission.  We obtained the FUSE spectra of NGC\,4631 from the MAST archive
(Program ID: P134; PI: E.  Murphy), and re-measured the \oVI\ emission.
Correcting for Milky Way extinction, we measure $5.2 \times
10^{-18}$\ergseccmarcsec\ (position A) and $6.7\times 10^{-18}$~\ergseccmarcsec\
(position B). Our observations are barely sensitive such surface brightnesses,
but at similar radial distances J1156 is significantly brighter in \oVI.
However NGC\,4631 forms stars at just $\sim 1$~\msunyr, which is $\sim 40$ times
lower than J1156 ($\sim 20$ times lower than \har), and over a significantly
more extended disk.  Thus we would very much expect the \oVI\ surface brightness
to be lower.

\section{Summary and Conclusions}\label{sect:conc}

We have used the \emph{Hubble Space Telescope} to image a star-forming galaxy in
nine filters between the far ultraviolet and $z-$band, enabling us to produce
high spatial resolution images in four emission lines: the \oVI\ doublet at
$\lambda\lambda=1032,1038$~\AA, and the \lya, \halpha, and \hbeta\ lines of \hI.
We detect \oVI\ in emission.  While not the first detection of \oVI\ emission
from a starburst galaxy, this is the first time the line emission has been
spatially resolved and mapped.  The \oVI\ emission takes the form of a very
extended halo; the surface brightness profile can be described by an exponential
function, with scale length of 7.5~kpc.  This is 10 times the scale length of
the nebular ionized gas measured by \halpha, and the continuous radiation from
massive stars (measured using the far UV continuum).  We detect \oVI\ emission
out to radii of 23~kpc.  Integrating the light profile to infinity we obtain the
first \oVI\ luminosity of a galaxy at any redshift:
$(20.5_{-1.2}^{+1.8})\times 10^{40}$~\ergsec.  

Ultraviolet spectroscopy obtained on the central pointing confirms \oVI\
emission, and also reveals the absorption component of the transition.  By
contrasting the absorbed flux with the halo emission we estimate that about 1/6
the total luminosity can arise (assuming spherical symmetry) due to stellar
continuum radiation being resonantly scattered by \oVI\ ions in the halo.  The
UV spectrum enables us to measure the column density of \oVI\ ions, which is
$\log(N_\mathrm{OVI}/\mathrm{cm}^{-2}) \approx 14.7$.  By assuming collisional
ionization equilibrium and the measured oxygen abundance, we can compute the
expected surface brightness in emission for a given physical thickness of the
\oVI\ region; equating this with the observed surface brightness we solve for
the size of the \oVI\ region along the line-of-sight, which is just 10~pc.  The
scale length of emission is 750 times this distance, implying the $T\approx
10^{5.5}$~K gas must be very clumpy, or, very roughly, fills just $\sim 10^{-3}$
of the volume.  This is consistent with -- and indeed expected of -- \oVI\ being
produced at the hydrodynamical interfaces of hotter outflowing gas where it
collides with and compresses against denser, cold circumgalactic gas
\citep[e.g.][]{Heckman2001ovi,Heckman2002}.  

From these calculations we derive the electron density in the coronal phase to
be 0.5~cm$^{-3}$.  A pressure of $P/k_\mathrm{B} \approx 1.6\times
10^5$~K~cm$^{-3}$ follows from the temperature and density, and we compare this
to the pressure of the \hII\ regions that is independently derived from the
[\sII] doublet in the SDSS spectrum.  The pressures in the two phases agree
within the $1-\sigma$ level.  Our calculations also include the derivation of
both the cooling time ($\approx 10^6$~yr) and the sound crossing time ($\approx
10^4$~yr).  From these two points we speculate that the pressure agreement may
be causal, and that both phases exist in pressure equilibrium with the hotter
phase that drives the outflow and fills the volume.  Indeed the disruption
timescale for the estimated cloud sizes (a few sound-crossing times) is so short
compared to the cooling time that this pressure confinement may be a requirement
to see the \oVI\ gas in emission at all. 

Assuming that the decreasing surface brightness with radius is due to a
geometrical decrease in the column length, and therefore the column density
(i.e. the volume density is constant), we compute the mass of \oVI\ ions to be
$3\times 10^{4}$~\msun.  Assuming the nebular metallicity is equivalent to that
of the \hII\ regions, we obtain a total mass of gas at $T\approx 300,000$~K to
be $5.5\times 10^{7}$~\msun.   This is just $\sim 2$~\% of the mass in the ionized ISM.

We calculate the current, instantaneous kinetic energy of the coronal phase of
the wind to be $5.9\times 10^{55}$~erg.  Comparably, $7.4\times10^{55}$~ergs
have been radiated in \oVI\ over the whole estimated lifetime of the starburst.
This however, is just $\sim 1$~\% of the $7.8\times10^{57}$~erg in mechanical
energy that has been returned to the ISM by the collective winds from massive
stars and the ejecta from core-collapse supernovae.  Thus even if the
thermalization efficiency is low, it is likely that the coronal phase contains a
small fraction of the available energy and mass at any given time.  Even over
the lifetime of the star formation episode, the radiative loss of wind energy
through the \oVI\ lines must be small. 

The \oVI\ absorbing gas, measured in the spectrum, is shown to be outflowing
with an average velocity of 380~\kms.  We compute the expected escape velocity,
which we determine to be significantly smaller than the speed of the outflowing
gas.  If the coronal gas is becoming unbound, then the metals entrained in the
outflow will no longer contribute to the metallicity of the galaxy, and will
instead enrich the local intergalactic medium.  However given the short cooling
time and cooling length of just 0.5~kpc, the gas will have cooled and recombined
before escaping.  These baryons will therefore become visible to Lyman series
absorption studies and the oxygen will be visible to \oI\ absorption (e.g. at
1302~\AA).

\acknowledgments

We are grateful to Emily Freeland, Joop Schaye, Arjan Bik, and Daniela Calzetti
for useful discussion regarding this, and followup work.  We thank the referee
for providing insightful feedback that greatly strengthened the manuscript.
M.H.  acknowledges the support of the Swedish Research Council
(Vetenskapsr{\aa}det) and the Swedish National Space Board (SNSB), and is Fellow
of the Knut and Alice Wallenberg Foundation.

{\it Facilities:} \facility{HST (ACS,WFC3,COS)}.

\bibliographystyle{apj}
\bibliography{a.bib}

\clearpage

\appendix

\section{Matching the Point Spread Function}\label{sect:psf}

\begin{figure}[h!]
\includegraphics[width=18cm]{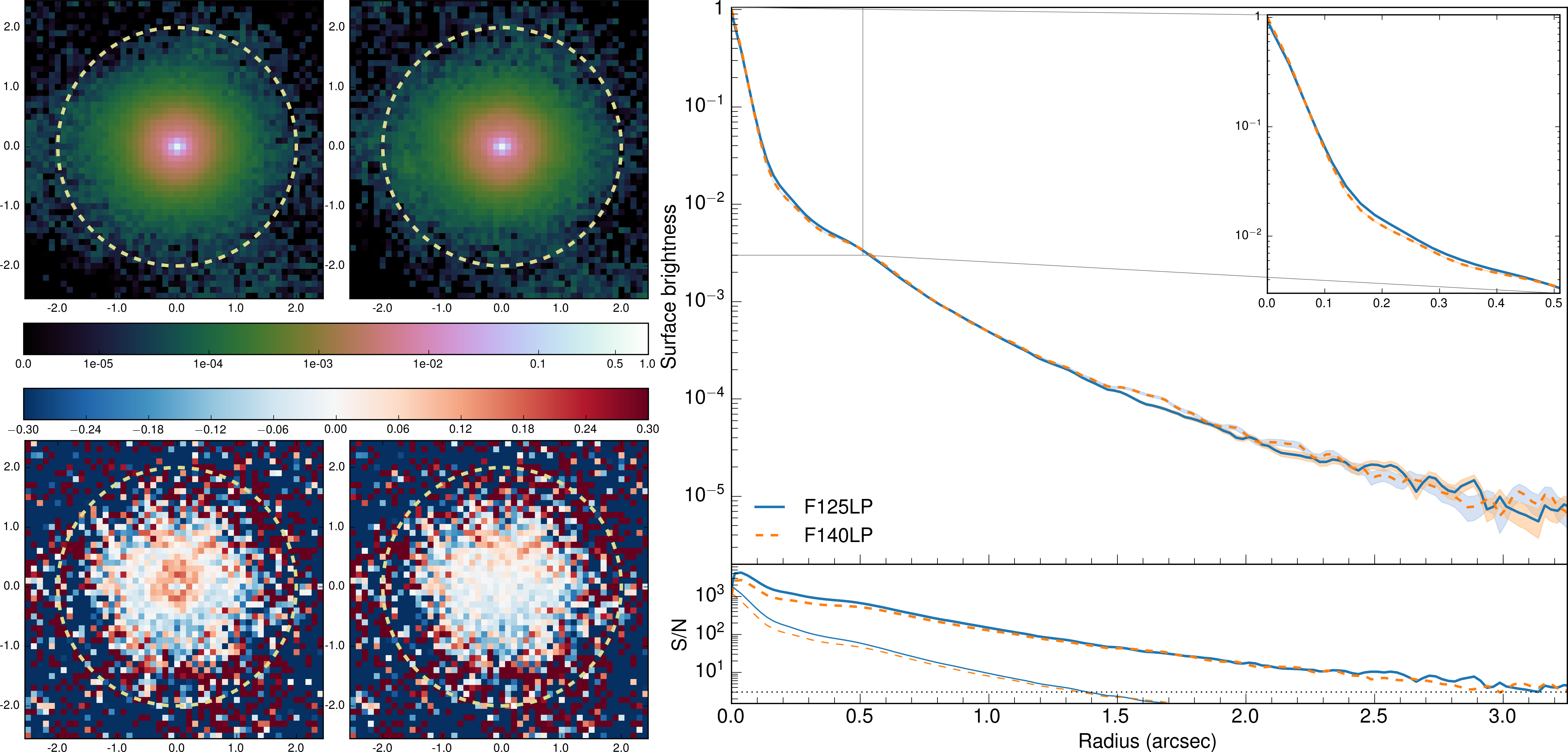}
\caption{The Point Spread Function of the SBC Filters. The top left panels show
the empirical PSF of the F125LP (left hand side) and F140LP (right hand side)
filters in log scale. The bottom left panels show the relative difference map,
(F125LP--F140LP) / F140LP, without any PSF matching on the left and with PSF
matching on the right. Note that the 2D PSF models are only trustworthy (at a
$3\sigma$ level) out to a radius of about 1.3 arcseconds. The top right panel
shows the surface brightness profiles of the two empirical PSFs with the shaded
regions showing the $1\sigma$ errors per annulus, and the inset showing the
inner 0.5 arcseconds of the profile. The lower right panel shows the
signal-to-noise $(S/N)$ of the surface brightness profile (thick lines), and the
average signal-to-noise measured in the 2D empirical PSFs within the same annuli
used to compute the surface brightness profile (thin lines).  }\label{fig:psf}
\end{figure}

The diffuse halo of \oVI\ emission could be mimicked if the F125LP Point Spread
Function (PSF) is broader than that of F140LP.  We used custom software to match
images to a common PSF. Point sources (stars) that are bright in the far UV are
rare, making empirical studies of the SBC PSFs challenging, and the models
available from STScI severely underestimate the power in the extended wings of
the PSF (at radii larger than ~1 arcsec). We instead build empirical PSF models
by examining multi-epoch calibration data of the globular cluster NGC\,6681.
These were obtained under many programs by the ACS instrument team at STScI,
stretching back to 2002 when the instrument was first mounted. We have obtained
all the calibration images taken in F125LP and F140LP from the MAST, drizzled
and stacked them using the same methods as for our science data. The total
exposure time in F125LP is 13,530 seconds, and 7,350 seconds in F140LP.  The
brightest star in NGC\,6681 has a FUV AB magnitude of 18.0, giving a countrate of
22 counts per second in the brightest pixel in F125LP.  This is just below the
non-linear regime of the SBC detector, indicating that calibration data are as
good as possible for deriving the PSF.  We then selected 45 (48) stars in the
F125LP (F140LP) images, extracted stamps, and stacked them to produce a very
deep point-source image. A model PSF consisting of a combination of three 2D
functions (an elliptical Moffat function, a radially symmetric Gaussian
function, and a symmetric fourth order polynomial function) was fitted to the
stacked image using least squares fitting with the LMFIT package for Python1.
The stacking was then redone iteratively with the contribution from nearby stars
removed by subtracting the model PSF. The final stacked images in the two
filters were normalized to have a total sum of 1 and are presented in
Figure~\ref{fig:psf}, while the final best fitting models were used to compute
the matching convolution kernel. 

The right panel of Figure~\ref{fig:psf} shows the extracted radial surface
brightness profiles of the empirical PSFs in F125LP and F140LP, demonstrating
that there is a small difference between the filters in the inner 0.5 arcseconds
of the PSF (shown in the inset), where the F125LP PSF is slightly more extended
than the F140LP PSF, but no difference within the uncertainties in the outer
parts of the profiles. The uncertainties for the radial profiles were estimated
by adding the sky noise, measured in the stacked images, to Poisson noise,
estimated from the PSF model, in quadrature. The azimuthally averaged profiles
have a S/N of at least 3 out to a radius of ~3 arcseconds.  Note that in
Figure~\ref{fig:radprof} the net \oVI\ emission begins significantly inside of
this radius. 

We then compute the relative difference between the filters as: 
\begin{equation}
\mathcal{R}(r) = \frac{\Sigma(f_{\mathrm{F125LP},i} -
f_{\mathrm{F140LP},i})}{\max(f_\mathrm{F140LP})} \bigg/ N
\end{equation}
where $f_\mathrm{F125LP}$ and $f_\mathrm{F140LP}$ designate the flux in the two
filters and summation is done over all pixels ($N$) inside a given radius $r$.
$\mathcal{R}$ is thus simply the difference in power per pixel at the given
radii divided by the peak count rate.  Within a radius of 0.5 arcseconds we find
a $\mathcal{R}$ value of $2\times 10^{-4}$ and within 3 arcseconds R is $3
\times 10^{-6}$. This clearly shows that the difference between the PSFs mainly
comes from the inner 0.5 arcseconds and not in the wings.

To account for the differences between the PSFs we compute convolution kernels
that matches all of the filters to a common PSF (see~\ref{sect:psfmatch}). The
lower part of the left panel in Figure~\ref{fig:psf} shows the relative
difference map for the non-matched and matched cases. While the regions outside
a radius of $\sim 1.3$ arcseconds (which is also seen in the $S/N$ plot in the
right panel) are too noisy for any meaningful comparisons to be made, the
matching shows a substantial improvement for the inner part of the PSF. The
$\mathcal{R}$ ($r=1$~arcsec) value for the matched case is $1\times 10^{-4}$,
roughly a factor of two smaller than in the non-matched case. At larger radius
the changes to $\mathcal{R}$ when matching is smaller because there is less
difference to correct.

The surface brightness profiles in the right panel of Figure~\ref{fig:psf} shows
that the PSF difference at a radius of $\sim 2$ arcseconds is extremely small
and in fact negative (there is more flux in the F140LP than the F125LP PSF at
that radius), with an $\mathcal{R}$ value of $-3\times 10^{-6}$ measured within
an annulus around $r=2$ arcseconds.  However, there are some ``bumps" in the
F140LP radial profile (also visible in the 2D image of the PSF as the asymmetric
emission at a radius $\sim 1.5$~arcseconds extending to the lower left) around
this radius that are unlikely to be part of the true profile. These bumps are
artifacts from stacking and gives rise to an additional systematic error (which
will only add flux to the PSF) in the profile, which leads us to conclude that
the difference at this radius is consistent with zero. Nevertheless, in our
final \oVI\ continuum subtraction we use PSF matched images that removes also this
difference. It should be noted that even without the PSF matching, the
difference at this radius is nowhere near enough to create the observed extended
emission we detect.

\clearpage

\end{document}